\documentclass[prd,preprint,tightenlines,floatfix,showpacs,preprintnumbers,nofootinbib,eqsecnum,superscriptaddress]{revtex4}

 \usepackage[dvips,final]{graphicx}
  \usepackage{amssymb}
   \usepackage{amsmath}
    \usepackage{amsfonts}
     \usepackage{epsfig}
      \usepackage{bm}

\newcommand{\bDelta}{\mbox{\boldmath $\Delta$}}

\newcommand{\bkappa}{\mbox{\boldmath $\kappa$}}
\newcommand{\bp}{\mbox{\boldmath $p$}}

\newcommand{\bk}{\mbox{\boldmath $k$}}

\newcommand{\bM}{\mbox{\boldmath $M$}}

\newcommand{\ket}[1]{| {#1} \rangle}

\newcommand{\half}{{1\over 2}}
\def\lsim{\mathrel{\rlap{\lower4pt\hbox{\hskip1pt$\sim$}}
    \raise1pt\hbox{$<$}}}         
\def\gsim{\mathrel{\rlap{\lower4pt\hbox{\hskip1pt$\sim$}}
    \raise1pt\hbox{$>$}}}         

\begin{document}

\title{Exclusive production of \mbox{\boldmath $\omega$} meson \\
in proton-proton collisions at high energies}

\author{Anna Cisek}
\email{anna.cisek@ifj.edu.pl}
\affiliation{Institute of Nuclear Physics PAN, PL-31-342 Cracow, Poland}
\author{Piotr Lebiedowicz}
\email{piotr.lebiedowicz@ifj.edu.pl}
\affiliation{Institute of Nuclear Physics PAN, PL-31-342 Cracow, Poland}
\author{Wolfgang Sch\"afer}
\email{wolfgang.schafer@ifj.edu.pl}
\affiliation{Institute of Nuclear Physics PAN, PL-31-342 Cracow, Poland}
\author{Antoni Szczurek}
\email{antoni.szczurek@ifj.edu.pl}
\affiliation{Institute of Nuclear Physics PAN, PL-31-342 Cracow, Poland}
\affiliation{University of Rzesz\'ow, PL-35-959 Rzesz\'ow, Poland}

\begin{abstract}
First we calculate cross section for the $\gamma p \to \omega p$
reaction from the threshold to very large energies.
At low energies the pion exchange is the dominant mechanism. 
At large energies the experimental cross section can be
well described within the $k_{t}$-factorization approach
by adjusting light-quark constituent mass.
Next we calculate differential distributions for the $p p \to p p \omega$
reaction at RHIC, Tevatron and LHC energies for the first time in the
literature. We consider photon-pomeron (pomeron-photon),
photon-pion (pion-photon) as well as diffractive hadronic bremsstrahlung mechanisms.
The latter are included in the meson/reggeon exchange picture with
parameters fixed from the known phenomenology.
Interesting rapidity distributions are predicted.
The hadronic bremsstrahlung contributions dominate at large (forward, backward)
rapidities. At small energies the photon-pomeron contribution
is negligible compared to the bremsstrahlung contributions.
It could be, however, easily identified
at large energies at midrapidities.
Absorptions effects are included and discussed.
Our predictions are ready for verification at RHIC and LHC.
\end{abstract}


\maketitle

\section{Introduction}

The mechanism of exclusive production of mesons at high energies became
recently a very active field of research (see \cite{ACF10} and references therein).
The recent works concentrated on the production of $\chi_c$ mesons
(see e.g. \cite{PST2008} and references therein) where 
the QCD mechanism is similar to the exclusive production of the Higgs boson. 
The latter process is an alternative to the inclusive production of 
the Higgs boson.
In the case of heavy vector quarkonia ($J/\Psi$, $\Upsilon$) the
dominant mechanism is photon-pomeron (pomeron-photon)
fusion (see e.g. \cite{SS07,RSS08}) which can be calculated
in the QCD language. 

The mechanism of exclusive light vector meson production was almost not 
studied in the literature, exception is $\phi$ meson \cite{CSS10}.
In the present paper we consider exclusive 
production of the isoscalar $\omega$ meson. This process was studied 
before only close to its production threshold.
Various theoretical models (see Refs. e.g.
\cite{Sibirstev, NSHHS98, Kaiser, Kaptari, NOHL07})
were developed to describe the experimental data (see Refs. \cite{low_data}).
Here the dominant mechanisms are meson exchange processes
as well as the $\omega$-meson bremsstrahlung driven by
meson exchanges.

How the situation changes at high-energy is interesting but was not
studied so far. While at low energy the meson exchanges ($\pi$, $\rho$,
$\omega$, $\sigma$) are the driving t-channel exchanges for the $\omega$
bremsstrahlung, at high energy their role is taken over by the pomeron 
exchange.
The latter will be treated here purely phenomenologically.
A similar hadronic bremsstrahlung-type mechanism 
is the Deck-mechanism for diffractive production
of $\pi N$ final states in $pp$ collisions \cite{Deck},
for a review, see e.g.\cite{Kaidalov}.

In the present paper we intend to make predictions for being in
operation colliders RHIC, Tevatron and LHC.
The hadronic bremsstrahlung mechanisms are expected to be enhanced for exclusive
production of $\omega$ meson compared to other vector mesons
as the $g_{\omega N N}$ coupling constant
is known to be large from low-energy phenomenology 
\cite{NOHL07,Bonn_potential}.
We will also show how important are the photoproduction mechanisms
discussed previously in the context of exclusive heavy vector quarkonium
production \cite{SS07,RSS08}.
The photoproduction mechanism constitutes a background for odderon-pomeron
exchanges possible in the discussed reaction. 
So far odderon exchange was discussed for the exclusive
$J/\Psi$ and $\Upsilon$ production \cite{odderon}.
The predicted by QCD odderon exchange was searched for in different
reactions.
No clear evidence was found so far. We shall comment on the issue
in the Result Section.

\section{Photoproduction mechanism for \boldmath $\gamma p \to \omega p$}

\subsection{Pomeron exchange}
\label{subsection:pomeron_exchange}

Let us concentrate on
the $\gamma p \to \omega p$ reaction which is a building block
for the $p p \to p p \omega$ reaction.
Photoproduction of the vector meson in photon-proton collisions is very 
interesting from both experimental and theoretical side.
The corresponding cross
sections have been measured 
by the ZEUS Collaboration at HERA at virtuality of photon
$Q^{2} \simeq 0$ GeV$^2$ for $\omega$ photoproduction \cite{ZEUS96}
and at large values $Q^{2}$ for $\omega$ electroproduction
$e p \to e \omega p$ \cite{ZEUS00}.
\begin{figure}[!h]
\includegraphics[width=0.6\textwidth]{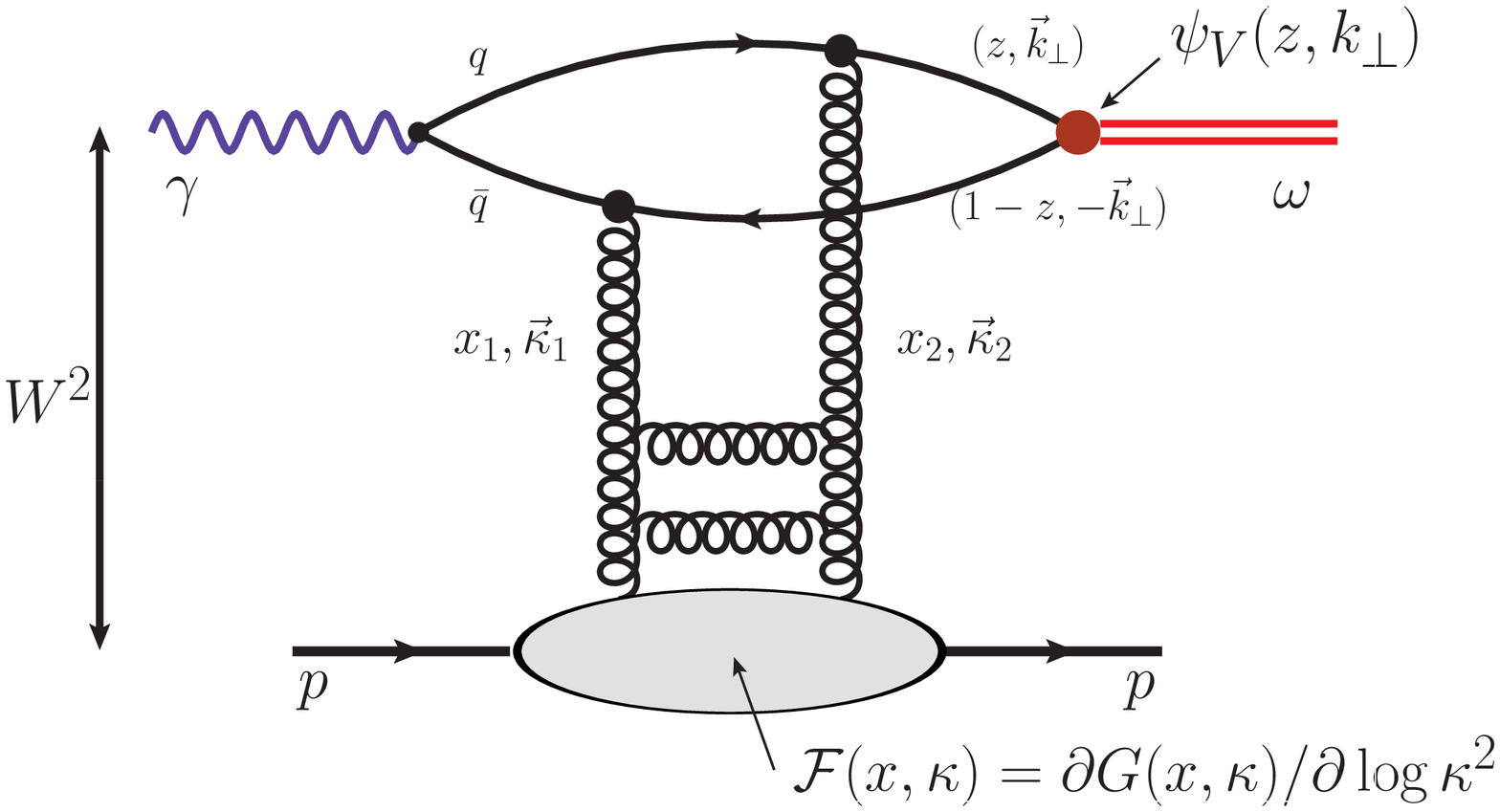}
   \caption{\label{VM_amplitude_photo}
   \small  A sketch of the amplitude for exclusive photoproduction 
           $\gamma p \to \omega p$ process.
           Some kinematical variables are shown in addition.}
\end{figure}
The amplitude
for this reaction is shown schematically 
in Fig.\ref{VM_amplitude_photo}.
The Pomeron exchange is modelled by a pQCD gluon ladder. 
The details how to calculate the amplitude are explained 
in Ref.\cite{INS06}.
The imaginary part of the amplitude for the $\gamma p \to \omega p$
process is written as:
\begin{eqnarray}
\Im m \, {\cal M}_{\lambda_{\gamma},\lambda_{V}}(W,t=-\bDelta^2, Q^{2}) &=&
W^2 \frac{c_{V} \sqrt{4 \pi \alpha_{em}}}{4 \pi^2}  \, 
\int \frac{d\bkappa^{2}}{\kappa^{4}}
\alpha_{S}(q^{2}) {\cal F}(x_{1},x_{2},\bkappa_{1},\bkappa_{2})\\ \nonumber
&\times& \int \frac{dz d^2\bk}{z(1-z)}
I_{\lambda_{\gamma},\lambda_{V}}(z, \bk, \bkappa_{1}, \bkappa_{2}, Q^{2}) \, ,
\end{eqnarray}
where 
the transverse momenta of gluons coupled to the $q\bar{q}$ pair
can be written as $\bkappa_{1}=\bkappa + \bDelta^{2}/2$ and 
$\bkappa_{2}=-\bkappa + \bDelta^{2}/2$,
$\bDelta^{2}$ is the (transverse) momentum transfer squared and
$\bk$ is the transverse momentum of the (anti-)quark.
The quantity ${\cal F}(x_{1},x_{2},\bkappa_{1},\bkappa_{2})$
is the off diagonal unintegrated gluon distribution.
Explicit expressions for $I_{\lambda_{\gamma},\lambda_{V}}$
can be found in Ref.\cite{INS06}.

In the forward scattering limit, i.e. for $\Delta^{2}=0$,
azimuthal integrations can be performed analytically.
The following representation for the imaginary part 
of the amplitude for the transverse polarization
for forward photoproduction $\gamma p \to \omega p$ process is used:

\begin{eqnarray}
\Im m \, {\cal M}(W,\Delta^2 = 0,Q^{2}=0) =
W^2 \frac{c_{V} \sqrt{4 \pi \alpha_{em}}}{4 \pi^2} \, 2 \, 
\int_0^1 \frac{dz}{z(1-z)} \nonumber
\int_0^\infty \pi d\bk^{2} \psi_V(z,\bk^{2})\\ 
\times \int_0^\infty {\pi d\bkappa^2 \over \bkappa^4} 
\alpha_S(q^2) {\cal{F}}(x_{eff},\bkappa^2)
\Big( A_0(z,\bk^2) \; W_0(\bk^2,\bkappa^2) 
    + A_1(z,\bk^2) \; W_1(\bk^2,\bkappa^2) \Big) \, ,
\end{eqnarray}
where
\begin{eqnarray}
A_0(z,\bk^2) &=& m_q^2 + \frac{\bk^2 m_q}{M + 2 m_q}  \, ,
\\
A_1(z,\bk^2) &=& \Big[ z^2 + (1-z)^2 
    - (2z-1)^2 \frac{m_q}{M + 2 m_q} \Big] \, \frac{\bk^2}{\bk^2+ \epsilon^{2}} \, ,
\\
W_0(\bk^2,\bkappa^2) &=& 
{1 \over \bk^2 + \epsilon^2} - {1 \over \sqrt{(\bk^2-\epsilon^2-\bkappa^2)^2 + 4 \epsilon^2 \bk^2}}
\, , 
\\
W_1(\bk^2,\bkappa^2) &=& 1 - { \bk^2 + \epsilon^2 \over 2 \bk^2}
\Big( 1 + {\bk^2 - \epsilon^2 - \bkappa^2 \over 
\sqrt{(\bk^2 - \epsilon^2 - \bkappa^2)^2 + 4 \epsilon^2 \bk^2 }}
\Big) \, ,
\end{eqnarray}
and $M$ is the invariant mass of the constituent $q\bar{q}$ system
\begin{eqnarray}
M=\frac{\bk^{2}+m_{q}^{2}}{z(1-z)}\, ,
\end{eqnarray}
where $z$ and $(1-z)$ are fractions of the longitudinal momentum of the $\omega$-meson
carried by a quark and antiquark, respectively.

The diagonal unintegrated gluon distribution 
can be emulated by taking the ordinary gluon distribution at
${\cal{F}}(x_{eff},\bkappa^2)$, where $x_{eff}=c_{skewed}(m_{\omega}^{2}/W^{2})$
with $c_{skewed}=0.41$ \cite{INS06}. The forward unintegrated gluon distribution
is taken from the work of Ivanov-Nikolaev \cite{IN02}, where it was found in the
analysis of the deep-inelastic scattering data.
The charge-isospin factor $c_{V}$ is
$c_{\omega}=1/\sqrt{2}(e_{u}+e_{d}) = 1/(3\sqrt{2})$.
In our calculation we choose the scale of the QCD running coupling
constant $\alpha_{S}$ at
$q^2 = \max \{ \bkappa^2, \bk^2 + m_q^2 \}$.

The full amplitude for the $\gamma p \to \omega p$ process is given as
\begin{eqnarray}
{\cal M}(W,\Delta^2,Q^{2}=0) = (i + \rho) \, \Im m {\cal M}(W,\Delta^2=0,Q^{2}=0)
\, \exp \Big( \frac{-B(W) \Delta^2}{2} \Big) \, ,
\label{full_amp}
\end{eqnarray}
where $\rho$ is a ratio of real to imaginary part of the amplitude and 
$B(W)$ is the slope parameter dependent on the
photon-proton center-of-mass energy and is parametrized as
\begin{eqnarray}
B(W) = B_0 + 2 \alpha'_{eff} \ln \Big( {W^2 \over W^2_0} \Big) \, ,
\end{eqnarray}
with: $W_0 = 95$ GeV, $B_{0} = 11$ GeV$^{-2}$, $\alpha'_{eff} = 0.25$ GeV$^{-2}$ \cite{H106}.

Our amplitude is normalizated to the total cross section:
\begin{eqnarray}
\sigma(\gamma p \to \omega p) = {1 +  \rho^2 \over 16 \pi B(W)} 
\Big| \Im m { {\cal M}(W,\Delta^2=0,Q^{2}=0) \over W^2 } \Big|^2 \, .
\end{eqnarray}

The radial light-cone wave function of the vector meson
can be regarded as a function of three-momentum 
$\bp = (\vec{p}, p_{z})$, where $\vec{p} = \vec{k}$,
$p_{z} = (2z-1)M/2$ then
\begin{eqnarray}
\psi_V(z,\vec{k}^{2}) \to \psi_V(p^{2}),\;
\frac{dz d^{2}\vec{k}}{z(1-z)} \to \frac{4d^{3}\bp}{M},\;
p^{2} = \frac{M^{2}-4m_{q}^{2}}{4}\,.
\end{eqnarray}
In our calculation we use a Gaussian wave function,
representing a standard harmonic-oscillator type quark model,
which turned out to be superior over a Coulomb wave function
(which a power-law tail in momentum space)
for $J/\Psi$, $\Upsilon$ and $\phi$ mesons 
exclusive photoproduction \cite{SS07,RSS08,CSS10}
\begin{eqnarray}
\psi_{V}(p^2) = N
\exp\left( - \frac{p^2 a_{1}^{2}}{2} \right) \, .  
\label{harmonic_oscillator_WF}
\end{eqnarray}
The parameter $a_1$ is obtained by fitting to the decay electronic width
%
\begin{eqnarray}
\Gamma (V \to e^+ e^-) = {4 \pi \alpha_{em}^2 c_{V}^{2} \over 3 m_{\omega}^3} \,
\cdot g_V^2\, ,
\end{eqnarray}
where $\Gamma (\omega \to e^+ e^-) = 0.6$ keV \cite{PDG}
and imposing the normalization condition
\begin{eqnarray}
1 = \frac{N_{c}4\pi}{(2\pi)^{3}}
\int_0^\infty p^{2}dp\, 4M \psi_{V}^{2}(p^2) \,.
\end{eqnarray}
In our calculation we use leading-order approximation, i.e.
we neglect a possible NLO $K$-factor. 
The parameter $g_V$ can be expressed in terms of the $\omega$-meson wave
function as \cite{INS06}
\begin{eqnarray}
g_V = {8 N_c \over 3} \int {d^3 \vec p \over (2 \pi)^3} 
(M + m_q) \,  \psi_V(p^2) \,.
\end{eqnarray}

\begin{figure}[!htb]
\begin{center}
\includegraphics[height=7.0cm]{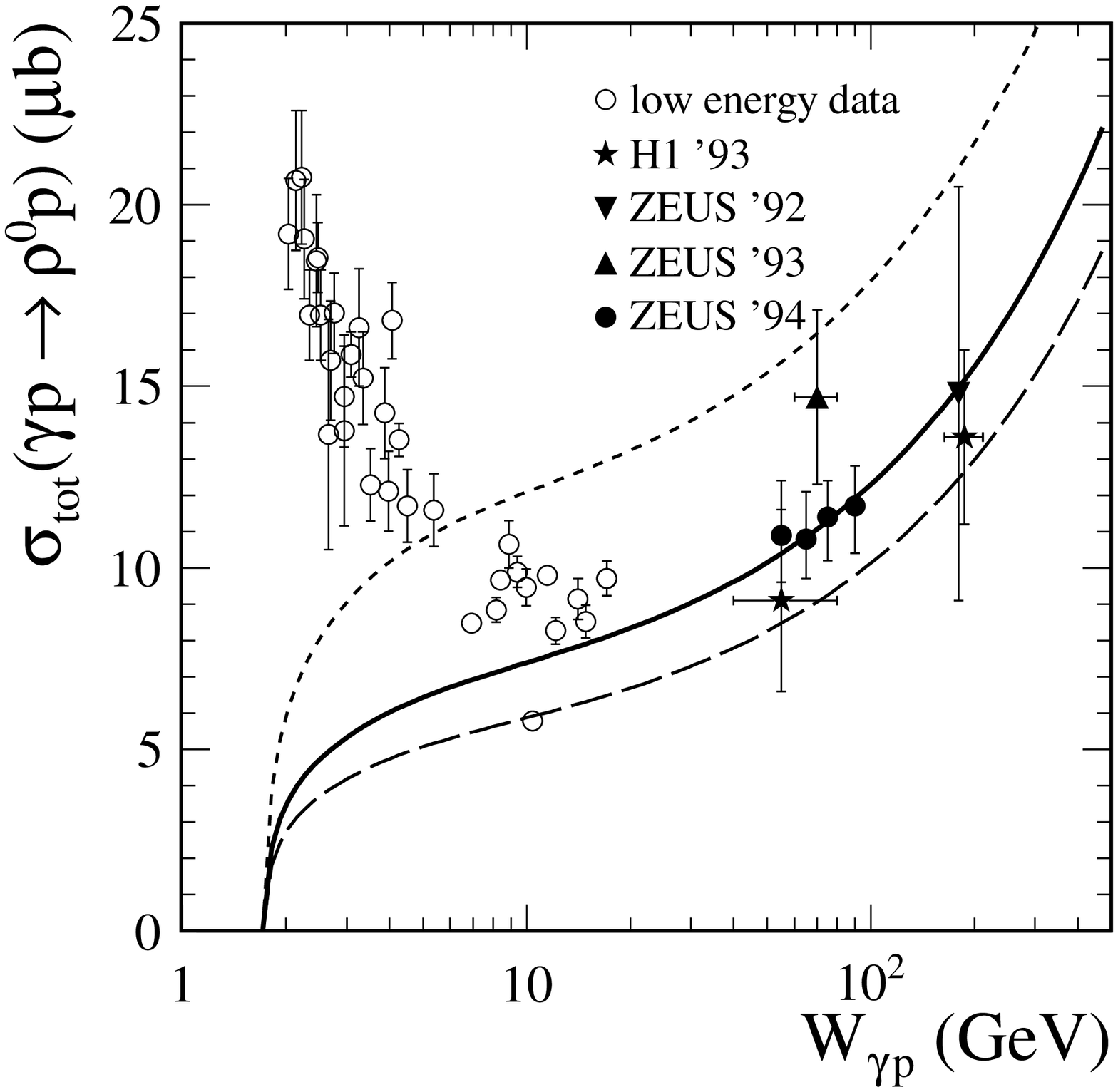}
\includegraphics[height=7.0cm]{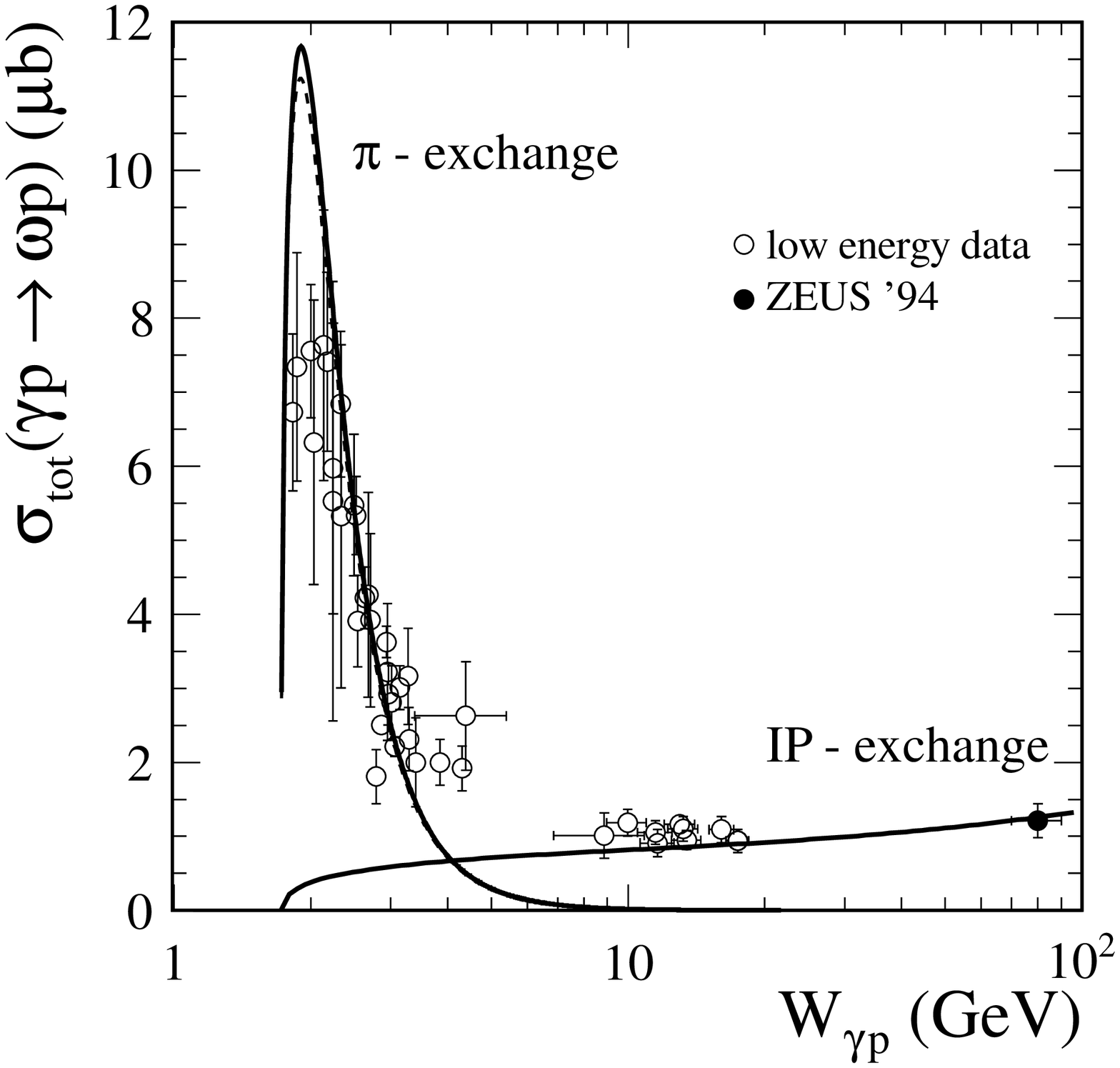}
\caption[*]{Total cross section for the photoproduction 
$\gamma p \to \rho^{0} p$ (left panel)
and $\gamma p \to \omega p$ (right panel) 
processes as a function of 
the photon-proton center-of-mass energy.
In the calculation 
of the $I\!\!P$-exchange mechanism
the Gaussian wave function
of the $\rho^{0}$ and $\omega$ mesons is used.
At low energies $\pi$-exchange is the dominant mechanism.
The curves are described in the text.
Our results are compared with the HERA data
\cite{ZEUS94,ZEUS95,ZEUS96,H196,ZEUS98} (solid marks) and with a compilation
of low energy data \cite{rho_lowenergy,omega_lowenergy} (open circles).}
\label{fig:w_sig}
\end{center}
\end{figure}

Having in view theoretical uncertainties in defining light quark mass
it is treated here as a model parameter.
In Fig.\ref{fig:w_sig} we show the total cross section for the exclusive 
$\gamma p \to \rho^{0} p$ (left panel) and $\gamma p \to \omega p$ (right
panel) processes as a function of the $\gamma p$ center-of-mass energy
$W_{\gamma p}$ for the photon virtuality $Q^2 = 0$ GeV$^{2}$.
Our results for exclusive $\rho^{0}$ and $\omega$ mesons production
are compared with the corresponding experimental data. 
For the $\rho^{0}$ meson we present results for three different values
of the $u$ and $d$ quark masses assumed here to be identical.
The dashed line (bottom)
is for $m_{q} = 0.33$ GeV, the dotted line (top) for $m_{q} = 0.22$ GeV and 
the thick solid line (fitted to experimental data) for $m_{q} = 0.3$ GeV.
Because the results for $m_{q} = 0.3$ GeV give the best description 
of experimental data, this mass will be used in further calculations.
In calculation the Gaussian wave function is used.
We see that it gives quite good description of the high-energy $\omega$-meson data.
At low energies the pion exchange 
mechanism dominates \cite{FS96,OTL01}.
This will be discussed in the following subsection. 

\subsection{Pion exchange}
\label{subsection:pion_exchange}
\begin{figure}[!h]
\includegraphics[width=0.3\textwidth]{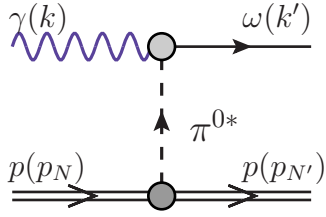}
   \caption{\label{pion_exchange}
   \small Diagram with the $\pi$-exchange for exclusive photoproduction 
          $\gamma p \to \omega p$.}
\end{figure}

The amplitude for the $\pi$-exchange can be written as:
\begin{eqnarray}
{\cal M}^{\pi^{0}-exch.}_{\lambda_{\gamma},\lambda_{N} \to \lambda_{\omega},\lambda_{N'}} &=&
g_{\omega \pi^{0} \gamma}\, F_{\omega \pi \gamma}(t)\,
\varepsilon^{\beta \mu \nu \lambda}\,
k_{\mu} \, k'_{\nu} \,
\varepsilon_{\beta}(k,\lambda_{\gamma})\,
\varepsilon^{*}_{\lambda}(k',\lambda_{\omega})\nonumber \\
&\times & g_{\pi^{0} NN}\, F_{\pi NN}(t)\, \dfrac{1}{t-m_{\pi}^{2}}\,
\bar{u}(p_{N'},\lambda_{N'}) \mathrm{i} \gamma_{5} u(p_{N},\lambda_{N})\,.
\label{pi_exchange}
\end{eqnarray}

The $g_{\omega \pi^0 \gamma}$ coupling constant in the formula above
is obtained from the $\omega$ partial decay width through the relation:
\begin{equation}
\Gamma(\omega \to \pi^0 \gamma) = {\cal BR}(\omega \to \pi^0 \gamma) \cdot \Gamma_{tot} =
\frac{g_{\omega \pi^0 \gamma}^2}{96 \pi}
\cdot m_{\omega}^3 \left( 1 - \frac{m_{\pi}^2}{m_{\omega}^2}  \right)^3
\; .
\end{equation}
Taking experimental partial decay width
$\Gamma(\omega \to \pi^0 \gamma)$ from
\cite{PDG} we get $g_{\omega \pi^0 \gamma} \approx$ 0.7 GeV$^{-1}$
which is consistent with the values used in Refs.\cite{OTL01,MNW10}
\footnote{Please note different normalization
convention of the coupling constant
in all the papers.}.
The pion-nucleon
coupling constant $g_{\pi NN}$ is relatively well known \cite{ELT02}.
In our calculations the coupling constant
$g_{\pi NN}^{2}/4\pi$ = 13.5.

We describe the low energy data 
shown in Fig.\ref{fig:w_sig} (right panel)
with $\Lambda_{mon} \approx$ 0.7 GeV
for the monopole form factors by the dashed line
\begin{equation}
F(t) = \frac{\Lambda_{mon}^2-m_{\pi}^{2}}{\Lambda_{mon}^2-t} \;
\end{equation}
or $\Lambda_{exp} \approx$ 0.8 GeV
for the exponential form factors by the solid line
\begin{equation}
F(t) = \exp \left( \frac{t - m_{\pi}^2}{\Lambda_{exp}^2}\right)\; .
\end{equation}
%
The cut-off parameters obtained from the fit are significantly smaller
than e.g. those used in the Bonn model \cite{Bonn_potential}.
Such soft form factors may be due to active coupling with
the $\pi N$ and $\rho N$ channels not included explicitly both here
nor in the literature. The pion exchange describes only angular
distributions at forward angles. At larger angles there are other
mechanisms as nucleon exchanges or s-channel nucleon resonances \cite{Laget00,OTL01}.
A more refined analysis in the peak region would require description
of new very precise CLAS Collaboration data \cite{CLAS09}
for full range angular distributions.
Such an analysis would need to include also channel couplings discussed
above.

The form factors found here will be used when
discussing $\gamma \pi^0$ and $\pi^0 \gamma$ exchanges
in the $p p \to p p \omega$ reaction.

\section{The amplitudes for the \boldmath $p p \to p p \omega$ reaction}

\subsection{\mbox{\boldmath $\gamma I\!\!P$} and
\mbox{\boldmath $I\!\!P \gamma$} exchanges}

\begin{figure}[!h]
\includegraphics[width=0.8\textwidth]{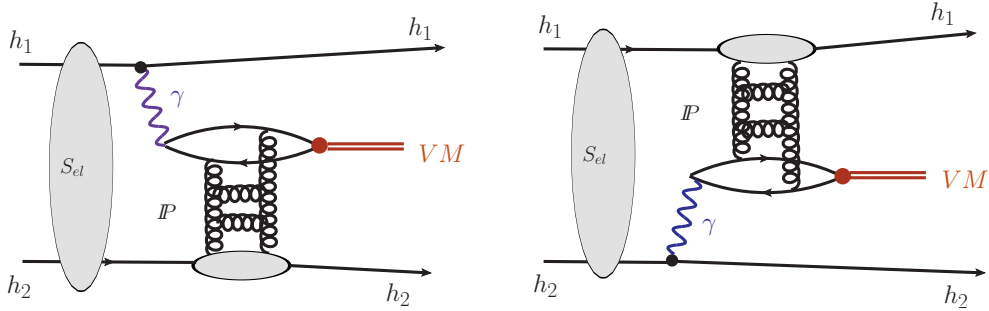}
   \caption{\label{VM_amplitude_abs}
   \small  A sketch of the exclusive photoproduction $p p \to p p \omega$ amplitudes 
with absorptive corrections.}
\end{figure}

The diagrams for the $p p$ and $p \bar p$ collisions in Fig.\ref{VM_amplitude_abs}
show schematically the amplitudes 
for photon-pomeron (pomeron-photon) exchanges
with absorptive correction, including
elastic rescattering.
The full amplitude (with absorptive correction)
for the $p p \to p \omega p$ or $p \bar p \to p \omega \bar p$
reactions can be written as
\begin{eqnarray}
\bM(\vec{p_1},\vec{p_2}) &=& \int{d^2 \vec{k} \over (2 \pi)^2} \,S_{el}(\vec{k}) \,\nonumber
\bM^{(0)}(\vec{p_1} - \vec{k}, \vec{p_2} + \vec{k})\\ 
&=& \bM^{(0)}(\vec{p_1},\vec{p_2}) - \delta \bM(\vec{p_1},\vec{p_2})\, ,
\label{amp_abs}
\end{eqnarray}
where
\begin{equation}
S_{el}(\vec{k}) = (2 \pi)^2 \delta^{(2)}(\vec{k}) - \half T(\vec{k}) \,,
\, \, \, T(\vec{k}) = \sigma^{pp}_{tot}(s) \, \exp \Big(-\half B_{el} \vec{k}^2 \Big) \, .
\end{equation}
Here $\vec{p_{1}}$ and $\vec{p_{2}}$ are the transverse momenta
of outgoing protons (RHIC, LHC) or proton and antiproton (Tevatron).
In practical evaluations we take $B_{el} = 14$ GeV$^{-2}$, $\sigma^{pp}_{tot} = 52$ mb
for the RHIC energy $W$ = 200 GeV, $B_{el} = 17$ GeV$^{-2}$, $\sigma^{p\bar{p}}_{tot} = 76$ mb
\cite{CDF94} for the Tevatron energy $W$ = 1.96 TeV and $B_{el} = 21$ GeV$^{-2}$,
$\sigma^{pp}_{tot} = 100$ mb for the LHC energy $W$ = 14 TeV.

The Born-amplitude (without absorptive correction) can be written
in the form of a two-dimensional vector
(corresponding to the two transverse (linear) polarizations
of the final state vector meson) \cite{SS07} as
\begin{eqnarray}
\bM^{(0)}(\vec{p_1},\vec{p_2}) &=& 
e_1 {2 \over z_1} {\vec{p_1} \over t_1} 
{\cal{F}}_{\lambda_1' \lambda_1}(\vec{p_1},t_1)\nonumber
{\cal {M}}_{\gamma^* h_2 \to V h_2}(s_2,t_2,Q_1^2)\\
&+& e_2 {2 \over z_2} {\vec{p_2} \over t_2} 
{\cal{F}}_{\lambda_2' \lambda_2}(\vec{p_2},t_2)
{\cal {M}}_{\gamma^* h_1 \to V h_1}(s_1,t_1,Q_2^2)\, ,
\end{eqnarray}
where 
${\cal {M}}_{\gamma^* h_2 \to V h_2}(s_2,t_2,Q_1^2)$ and 
${\cal {M}}_{\gamma^* h_1 \to V h_1}(s_1,t_1,Q_2^2)$ are the amplitudes
for photoproduction discussed above (see (\ref{full_amp})).
Because of the presence of the Dirac electromagnetic form factor
of the proton/antiproton
only small $Q_{1}^{2}$ and $Q_{2}^{2}$
enter the amplitude for the hadronic process.
This means that in practice one can put $Q_1^2=Q_2^2=0$ GeV$^{2}$
for the $\gamma^* p \to V p$ amplitudes.
We used the assumption of s-channel helicity conservation
in the $\gamma \to \omega$ transition,
$\lambda_{\gamma}=\lambda_{V}$.

%
%

The absorptive correction for the amplitude have the form:
\begin{eqnarray}
\delta \bM(\vec{p_1},\vec{p_2}) = \int {d^2\vec{k} \over 2 (2\pi)^2} \, T(\vec{k}) \,
\bM^{(0)}(\vec{p_1} -\vec{k},\vec{p_2} +\vec{k}) \, .
\end{eqnarray}

The differential cross section is expressed 
in terms of the amplitude $\bM$ as
\begin{eqnarray}
d \sigma = { 1 \over 512 \pi^4 s^2 } | \bM |^2 \, dy dt_1 dt_2
d\phi \, .
\end{eqnarray}
where $y$ is rapidity of the $\omega$ meson,
$t_{1,2}\simeq-\vec{p}^{\; 2}_{1,2}$ and $\phi$
is the azimuthal angle between 
transverse momenta $\vec{p_{1}}$ and $\vec{p_{2}}$.
\footnote{In the following for brevity we shall
use notation $t_{1,2}$ which means $t_{1}$ or $t_{2}$.}

\subsection{\mbox{\boldmath $\gamma \pi^{0}$} and
\mbox{\boldmath $\pi^{0} \gamma$} exchanges}

\begin{figure}[!h]
\includegraphics[width=4.5cm]{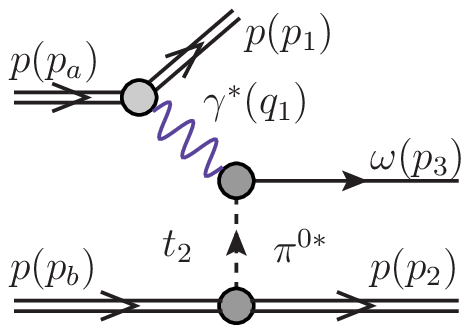}
\includegraphics[width=4.5cm]{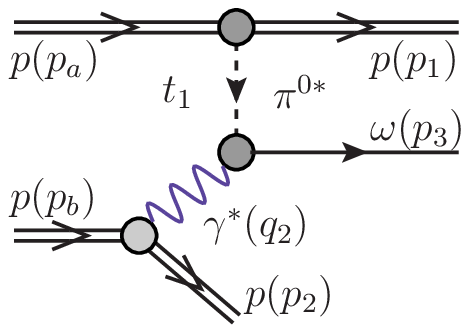}
   \caption{\label{fig:photon-pion_diagrams}
   \small  Diagrams with the $\gamma \pi^{0}$ and  
           $\pi^{0} \gamma$ exchange amplitudes in the $p p \to p p \omega$ reaction.
}
\end{figure}

As shown in Fig.\ref{fig:w_sig} the QCD mechanism
discussed in subsection \ref{subsection:pomeron_exchange} 
does not describe
the huge close-to-threshold enhancement 
of the cross section.
This indicates a presence of another mechanisms of omega photoproduction. 
Neutral pion exchange is the best candidate which describes
the low energy data as discussed
in subsection \ref{subsection:pion_exchange}.
Therefore for the $p p \to p p \omega$ reaction we should
include also photon-pion and pion-photon exchanges.
The underlying mechanisms are shown in Fig.\ref{fig:photon-pion_diagrams}.

The amplitudes for the two new processes can be easily written as:
\begin{eqnarray}
{\cal M}^{\gamma \pi^{0}-exch.}_{\lambda_{a}\lambda_{b} \to \lambda_{1}\lambda_{2}\lambda_{3}} &=&
e\, F_{1}(t_{1}) \,
\bar{u}(p_{1},\lambda_{1}) \gamma^{\alpha} u(p_{a},\lambda_{a})\nonumber \\
&\times &
\dfrac{-g_{\alpha \beta}}{t_{1}}\,
g_{\omega \pi^{0} \gamma} F_{\gamma \pi \to \omega}(t_1, t_2) \; 
\varepsilon^{\beta \mu \nu \lambda}\,
q_{1 \mu} \, p_{3 \nu}
\varepsilon_{\lambda}^{*}(p_{3},\lambda_{3})\nonumber \\
&\times &
g_{\pi^{0} NN} F_{\pi NN}(t_{2})\,
\dfrac{1}{t_{2}-m_{\pi}^{2}} \,
 \bar{u}(p_{2},\lambda_{2}) \mathrm{i} \gamma_{5}
 u(p_{b},\lambda_{b})\,,\\
%
{\cal M}^{\pi^{0} \gamma -exch.}_{\lambda_{a}\lambda_{b} \to \lambda_{1}\lambda_{2}\lambda_{3}} &=&
g_{\pi^{0} NN} F_{\pi NN}(t_{1}) \,
\dfrac{1}{t_{1}-m_{\pi}^{2}} \,
\bar{u}(p_{1},\lambda_{1}) \mathrm{i} \gamma_{5} u(p_{a},\lambda_{a})\nonumber \\
&\times &
\dfrac{-g_{\alpha \beta}}{t_{2}}\,
g_{\omega \pi^{0} \gamma} F_{\gamma \pi \to \omega}(t_2, t_1) \; 
\varepsilon^{\beta \mu \nu \lambda}\,
q_{2 \mu} \, p_{3 \nu}
\varepsilon_{\lambda}^{*}(p_{3},\lambda_{3})\nonumber \\
&\times &
e\, F_{1}(t_{2}) \,
\bar{u}(p_{2},\lambda_{2}) \gamma^{\alpha} u(p_{b},\lambda_{b})\,,
\label{pigam_amp}
\end{eqnarray}
where $F_{1}(t_{1,2})$ are the Dirac electromagnetic form factors of 
participating protons.
The $g_{\omega \pi^0 \gamma}$ constant was obtained
from the omega partial decay width as discussed in subsection \ref{subsection:pion_exchange}.
The coupling of the pion to the nucleon is described by the pion-nucleon
coupling constant $g_{\pi NN}$ and the corresponding form factor is taken
in the exponential form:
\begin{equation}
F_{\pi NN}(t_{1,2}) = \exp \left( \frac{t_{1,2} - m_{\pi}^2}{\Lambda_{\pi NN}^2}
\right)
\, .
\label{piNN_ff}
\end{equation}
The central vertices involve off-shell particles. Here the
$\gamma \pi^0$ and $\pi^0 \gamma$ form factors are taken in the
following factorized form:
\begin{eqnarray}
F_{\gamma \pi \to \omega}(t_1, t_2) &=& 
\frac{m_{\rho}^2}{m_{\rho}^2 - t_1}
\exp \left( \frac{ t_2 - m_{\pi}^2 }{\Lambda_{\omega \pi \gamma}^2}
\right)
\, .
\label{omega_pi_gamma_ff}
\end{eqnarray}
The factor describing the virtual photon coupling
is taken as in the vector dominance model.
In practical calculations we take:
$\Lambda_{\pi NN}$ = 0.8 GeV and $\Lambda_{\omega \pi \gamma}$ = 0.8 GeV
as found from the fit to the $\gamma p \to \omega p$
experimental data.

At high-energies often light-cone form factors are used instead
of the $t_1$ or $t_2$ dependent ones discussed above (see Eq.(\ref{piNN_ff})).
In such an approach
the pion is rather a constituent of the initial proton.
Then the form factors are parametrized in terms of the squared invariant masses
of the $\pi N$ system:
\begin{eqnarray}
M_{2.\pi N}^2(z_{2},p_{2t}^2)
&=& \frac{m_N^2 + p_{2t}^2}{z_2}
            + \frac{m_{\pi}^2 + p_{2t}^2}{1 - z_2}\, , \nonumber \\
M_{1,\pi N}^2(z_{1},p_{1t}^2)
&=& \frac{m_N^2 + p_{1t}^2}{z_1}
            + \frac{m_{\pi}^2 + p_{1t}^2}{1 - z_1}\, ,
\end{eqnarray}
where the longitudinal momentum fractions of outgoing protons with respect
to the initial protons can be calculated from energies and $z$-components 
of momenta of participating protons
\begin{eqnarray}
z_2 &=& (p_{20} - p_{2z}) / (p_{b0} - p_{bz}) \, , \nonumber \\
z_1 &=& (p_{10} + p_{1z}) / (p_{a0} + p_{az}) \, .
\end{eqnarray}
The light-cone form factors are parametrized then as
\begin{eqnarray}
F_{\pi NN}(M_{2,\pi N}^2)
&=& \exp\left( -\frac{ M_{2,\pi N}^2(z_{2},p_{2t}^2) - 
m_N^2}{2 \Lambda_{LC}^2}
\right)
\, , \nonumber \\
F_{\pi NN}(M_{1,\pi N}^2) 
&=& \exp\left( -\frac{ M_{1,\pi N}^2(z_{1},p_{1t}^2) - 
m_N^2}{2 \Lambda_{LC}^2} \right)
\, .
\label{light-cone-ff} 
\end{eqnarray}
The parameter $\Lambda_{LC}$ in the light-cone parametrization
was fitted in Ref.\cite{HSS96}
to the data on forward nucleon production and
the value $\Lambda_{LC} = 1.1$ GeV was found.

The amplitude for processes shown in Fig.\ref{fig:photon-pion_diagrams}
is calculated numerically for each point in the phase space.
In calculating cross section we perform integration in
$\log_{10}(p_{1t})$ (for $\gamma \pi$-exchange)
and $\log_{10}(p_{2t})$ (for $\pi \gamma$-exchange)
instead in $p_{1t}$ and $p_{2t}$.

\section{Hadronic bremsstrahlung mechanisms}

\subsection{The amplitude in the standard approach}

\begin{figure}[!h]
a)\includegraphics[width=4.5cm]{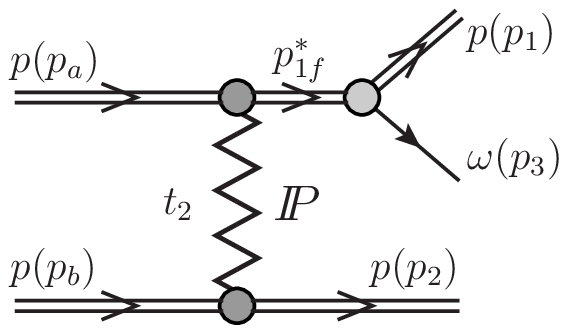}
c)\includegraphics[width=4cm]{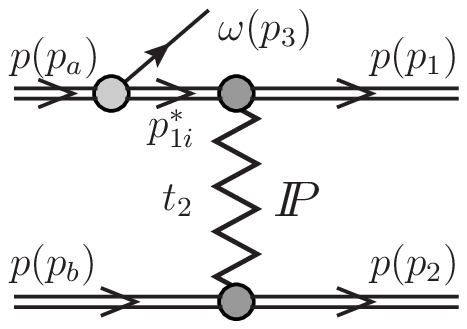}
e)\includegraphics[width=4cm]{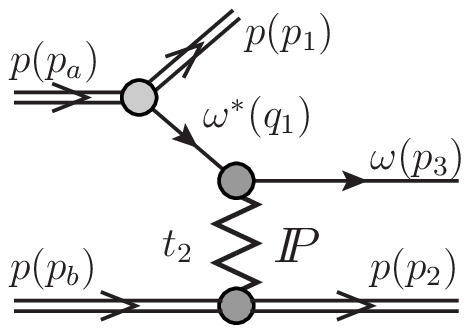} \\
b)\includegraphics[width=4.5cm]{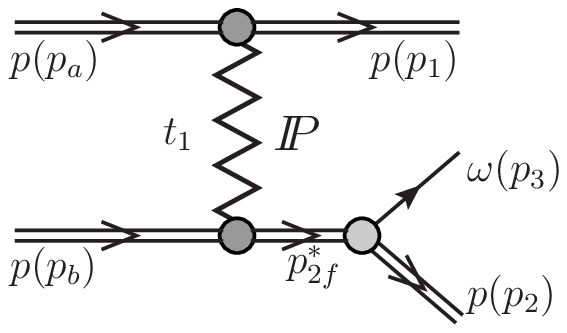}
d)\includegraphics[width=4cm]{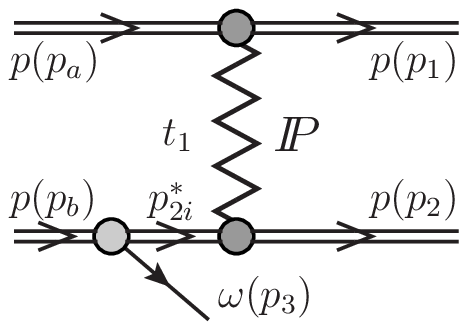}
f)\includegraphics[width=4cm]{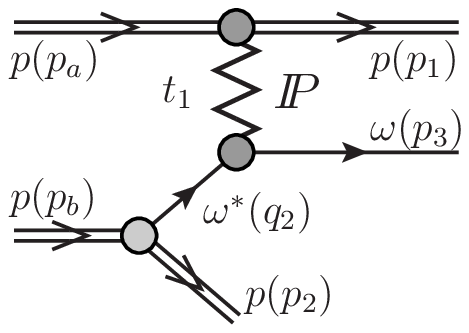}
   \caption{\label{fig:bremsstrahlung_diagrams}
   \small  Diagrams of the hadronic bremsstrahlung amplitudes
included in the present paper.
}
\end{figure}

The strong coupling of the $\omega$ meson to the nucleon causes
that the hadronic bremsstrahlung mechanisms become important.
The bremsstrahlung mechanisms for exclusive production of $\omega$
discussed here are shown schematically in Fig.\ref{fig:bremsstrahlung_diagrams}.
In the case of $\omega$ production the diagrams with intermediate
nucleon resonances are negligible (see \cite{PDG}). 
Because at high energy the pomeron is the driving mechanism
of bremsstrahlung it is logical to call the mechanisms
diffractive bremsstrahlung to distinguish from the low-energy
bremsstrahlung driven by meson exchanges.

It is straightforward to evaluate the contribution
of diagrams shown in Fig.\ref{fig:bremsstrahlung_diagrams}.
The Born amplitudes read:
\begin{eqnarray}
{\cal M}^{(a)}_{\lambda_{a}\lambda_{b} \to \lambda_{1}\lambda_{2}\lambda_{3}} &=&
\bar{u}(p_{1},\lambda_{1}) \varepsilon_{\mu}^*(p_{3},\lambda_{3}) \gamma^{\mu}  S_{N}(p_{1f}^*) u(p_{a},\lambda_{a})\;
g_{\omega NN}\; 
F_{\omega N^{*}N}(p_{1f}^{*2}) \; F_{I\!\!P NN^{*}}(p_{1f}^{*2}) \nonumber \\
&\times &
\mathrm{i} s_{ab} C_{I\!\!P}^{NN} \left( \frac{s_{ab}}{s_{0}}\right)^{\alpha_{I\!\!P}(t_{2})-1} 
\exp\left(\frac{B_{I\!\!P}^{NN} t_{2}}{2}\right)\;
\delta_{\lambda_{2}\lambda_{b}},
\label{brem_a}
\end{eqnarray}
\begin{eqnarray}
{\cal M}^{(b)}_{\lambda_{a}\lambda_{b} \to \lambda_{1}\lambda_{2}\lambda_{3}} &=&
\bar{u}(p_{2},\lambda_{2}) \varepsilon_{\mu}^*(p_{3},\lambda_{3}) \gamma^{\mu}  S_{N}(p_{2f}^{*2}) u(p_{b},\lambda_{b})\;
g_{\omega NN}\;
F_{\omega N^{*}N}(p_{2f}^{*2}) \; F_{I\!\!P NN^{*}}(p_{2f}^{*2}) \nonumber \\
&\times &
\mathrm{i} s_{ab} C_{I\!\!P}^{NN} \left( \frac{s_{ab}}{s_{0}}\right)^{\alpha_{I\!\!P}(t_{1})-1}
\exp\left(\frac{B_{I\!\!P}^{NN} t_{1}}{2}\right)\;
\delta_{\lambda_{1}\lambda_{a}},
\label{brem_b}
\end{eqnarray}
\begin{eqnarray}
{\cal M}^{(c)}_{\lambda_{a}\lambda_{b} \to \lambda_{1}\lambda_{2}\lambda_{3}} &=&
\bar{u}(p_{1},\lambda_{1})S_{N}(p_{1i}^{*2}) \varepsilon_{\mu}^*(p_{3},\lambda_{3}) \gamma^{\mu} u(p_{a},\lambda_{a})\;
g_{\omega NN}\;
F_{\omega NN^{*}}(p_{1i}^{*2}) \; F_{I\!\!P N^{*}N}(p_{1i}^{*2}) \nonumber \\
&\times &
\mathrm{i} s_{12} C_{I\!\!P}^{NN} \left( \frac{s_{12}}{s_{0}}\right)^{\alpha_{I\!\!P}(t_{2})-1} 
\left( \frac{s_{13}}{s_{th}}\right)^{\alpha_{N}(p_{1i}^{*2})-\frac{1}{2}}\;
\exp\left(\frac{B_{I\!\!P}^{NN} t_{2}}{2}\right)\;
\delta_{\lambda_{2}\lambda_{b}},
\label{brem_c}
\end{eqnarray}
\begin{eqnarray}
{\cal M}^{(d)}_{\lambda_{a}\lambda_{b} \to \lambda_{1}\lambda_{2}\lambda_{3}} &=&
\bar{u}(p_{2},\lambda_{2})S_{N}(p_{2i}^{*2}) \varepsilon_{\mu}^*(p_{3},\lambda_{3}) \gamma^{\mu} u(p_{b},\lambda_{b})\;
g_{\omega NN}\;
F_{\omega NN^{*}}(p_{2i}^{*2}) \; F_{I\!\!P N^{*}N}(p_{2i}^{*2}) \nonumber \\
&\times &
\mathrm{i} s_{12} C_{I\!\!P}^{NN} \left( \frac{s_{12}}{s_{0}}\right)^{\alpha_{I\!\!P}(t_{1})-1}
\left( \frac{s_{23}}{s_{th}}\right)^{\alpha_{N}(p_{2i}^{*2})-\frac{1}{2}}\;
\exp\left(\frac{B_{I\!\!P}^{NN} t_{1}}{2}\right)\;
\delta_{\lambda_{1}\lambda_{a}}.
\label{brem_d}
\end{eqnarray}
The diagrams for the interaction with emitted $\omega$:
\begin{eqnarray}
{\cal M}^{(e)}_{\lambda_{a}\lambda_{b} \to \lambda_{1}\lambda_{2}\lambda_{3}} &=&
\bar{u}(p_{1},\lambda_{1}) \gamma^{\mu} u(p_{a},\lambda_{a}) 
S_{\mu \nu}(t_{1}) \varepsilon^{\nu*}(p_{3},\lambda_{3})\;
g_{\omega NN}\;  F_{\omega^{*}NN}(t_{1}) F_{I\!\!P \omega^{*}\omega}(t_{1})\; \nonumber \\
&\times &
\mathrm{i} s_{23} C_{I\!\!P}^{\omega N} \left( \frac{s_{23}}{s_{0}}\right)^{\alpha_{I\!\!P}(t_{2})-1}\;
\left( \frac{s_{13}}{s_{th}}\right)^{\alpha_{\omega}(t_{1})-1}\;
\exp\left(\frac{B_{I\!\!P}^{\omega N} t_{2}}{2}\right)\; 
\delta_{\lambda_{2}\lambda_{b}},
\label{brem_e}
\end{eqnarray}
\begin{eqnarray}
{\cal M}^{(f)}_{\lambda_{a}\lambda_{b} \to \lambda_{1}\lambda_{2}\lambda_{3}} &=&
\bar{u}(p_{2},\lambda_{2}) \gamma^{\mu} u(p_{b},\lambda_{b}) 
S_{\mu \nu}(t_{2})  \varepsilon^{\nu*}(p_{3},\lambda_{3})\;
g_{\omega NN}\; F_{\omega^{*}NN}(t_{2}) F_{I\!\!P \omega^{*}\omega}(t_{2})\; \nonumber \\
&\times &
\mathrm{i} s_{13} C_{I\!\!P}^{\omega N} \left( \frac{s_{13}}{s_{0}}\right)^{\alpha_{I\!\!P}(t_{1})-1}\; 
\left( \frac{s_{23}}{s_{th}}\right)^{\alpha_{\omega}(t_{2})-1}\;
\exp\left(\frac{B_{I\!\!P}^{\omega N} t_{1}}{2}\right)\;
\delta_{\lambda_{1}\lambda_{a}},
\label{brem_f}
\end{eqnarray}
where $s_{0} = 1$ GeV$^2$ and $s_{th} = (m_{N}+m_{\omega})^2$.

In the above equations
$u(p_{i},\lambda_{i})$, $\bar{u}(p_{f},\lambda_{f})=u^{\dagger}(p_{f},\lambda_{f})\gamma^{0}$
are the Dirac spinors (normalized as $\bar{u}(p) u(p) = 2 m_{N}$) of 
the initial and outgoing protons with the four-momentum $p$ and 
the helicities $\lambda$.
The propagators of nucleons and $\omega$ meson can be written as
\begin{eqnarray}
S_{N}(p_{1f,2f}^{*2}) &=& {\frac{\mathrm{i}(p^{*}_{1f,2f_{\nu}} \gamma^{\nu} + m_{N})}{p_{1f,2f}^{*2} - m_{N}^{2}}}\,,\nonumber \\
S_{N}(p_{1i,2i}^{*2}) &=& {\frac{\mathrm{i}(p^{*}_{1i,2i_{\nu}} \gamma^{\nu} + m_{N})}{p_{1i,2i}^{*2} - m_{N}^{2}}}\,,\nonumber \\
S_{\mu \nu}(t) &=& \frac{-g_{\mu \nu}+\frac{q_{\mu}q_{\nu}}{m_{\omega}^{2}}}
{t - m_{\omega}^{2}}\,,
\label{propagators}
\end{eqnarray}
where
$t_{1,2}=(p_{a,b}-p_{1,2})^{2}=q_{1,2}^{2}$,
$p_{1i,2i}^{*2}=(p_{a,b}-p_{3})^{2}$,
$p_{1f,2f}^{*2}=(p_{1,2}+p_{3})^{2}$
are the four-momenta squared of objects in the middle of diagrams and
$s_{ij}=(p_{i}+p_{j})^{2}$ are squared invariant masses of the 
$(i,j)$ system.

The factor $g_{\omega NN}$ is the omega nucleons coupling constant.
Different values have been used in the literature \cite{Bonn_potential}.
In our calculations the coupling constant is taken as
$g^2_{\omega N N}/4\pi = 10$. Similar value was used
in Refs.\cite{Kaiser,NOHL07}.

Using the known strength parameters for the $NN$ and $\pi N$ scattering
fitted to the corresponding total cross sections
(the Donnachie-Landshoff model \cite{DL92}) we obtain
$C_{I\!\!P}^{NN} = 21.7$ mb and
$C_{I\!\!P}^{\omega N} = C_{I\!\!P}^{\pi N} = 13.63$ mb.
The pomeron/reggeon trajectory determined from elastic and total cross sections
is taken in the linear approximation in $t$
($\alpha(t)=\alpha(0)+\alpha' \, t$)
\begin{eqnarray}
\alpha_{I\!\!P}(t) = 1.0808 + 0.25\, t\,,
\qquad \alpha_{\omega}(t) = 0.5 + 0.9\, t \, ,
\label{trajectories}
\end{eqnarray}
where the values of the intercept $\alpha(0)$ and
the slope of the trajectory $\alpha'$ are also
taken from the Donnachie-Landshoff model \cite{DL92} for consistency.
The slope parameter can be written as
\begin{eqnarray}
B(s) = B_{0} + 2 \alpha'_{I\!\!P} \ln \left( \frac{s}{s_{0}}\right)\,.
\label{slope}
\end{eqnarray}
In our calculation we use $B_{0}$:
$B^{\omega N}_{I\!\!P} = 5.5$ GeV$^{-2}$ and $B^{NN}_{I\!\!P} = 9$ GeV$^{-2}$.

The extra factors $F_{\omega NN}$ and $F_{I\!\!P NN}$
(or $F_{I\!\!P \omega \omega}$)
allow for modification when one of the nucleons
or the $\omega$-meson is off its mass shell.
We parametrize all the form factors 
in the following exponential form:
%
\begin{eqnarray}
F_{\omega NN}(p_{1f,2f}^{*2})&=&\exp\left(\frac{-(p_{1f,2f}^{*2}-m_{N}^{2})}
{\Lambda^{2}}\right),
F_{I\!\!P NN}(p_{1f,2f}^{*2})=\exp\left(\frac{-(p_{1f,2f}^{*2}-m_{N}^{2})}
{\Lambda_{I\!\!P NN}^{2}}\right),\nonumber \\
F_{\omega NN}(p_{1i,2i}^{*2})&=&\exp\left(\frac{p_{1i,2i}^{*2}-m_{N}^{2}}
{\Lambda^{2}}\right),
F_{I\!\!P NN}(p_{1i,2i}^{*2})=\exp\left(\frac{p_{1i,2i}^{*2}-m_{N}^{2}}
{\Lambda_{I\!\!P NN}^{2}}\right),\nonumber \\
F_{\omega NN}(t_{1,2})&=&\exp\left(\frac{t_{1,2}-m_{\omega}^{2}}
{\Lambda^{2}}\right),
F_{I\!\!P \omega \omega}(t_{1,2})=\exp\left(\frac{t_{1,2}-m_{\omega}^{2}}
{\Lambda_{I\!\!P \omega \omega}^{2}}\right).
\label{exp_form_factors}
\end{eqnarray}
In general, the cut-off parameters are not known
but could be fitted to the
(normalized) experimental data.
From our general experience in hadronic physics
we expect $\Lambda \approx \Lambda_{I\!\!P NN}\approx \Lambda_{I\!\!P \omega \omega} = 1$ GeV.
We shall discuss how the uncertainties
of the form factors influence our final results.

Since the amplitudes given by formulas (\ref{brem_e}, \ref{brem_f}) are as if for 
$\omega$ meson exchanges they are corrected by the factors
$\left( \frac{s_{i3}}{s_{th}} \right)^{\alpha_{\omega}(t_{1,2})-1}$
to reproduce the high-energy Regge dependence.
We improve also the parametrization of the amplitudes
(\ref{brem_c}, \ref{brem_d}) by the factors
$\left( \frac{s_{i3}}{s_{th}} \right)^{\alpha_{N}(p^{*2}_{1i,2i})-\frac{1}{2}}$,
where the degenerate nucleon trajectory is
$\alpha_{N}(p^{*2}_{1i,2i})=-0.3 + \alpha'_{N} \, p^{*2}_{1i,2i}$, with $\alpha'_{N}=0.9$ GeV$^{-2}$.

We have chosen a representation for the polarization vectors of 
the $\omega$-meson in the helicity states $\lambda_{3} = 0,\pm 1$. 
The polarization vectors are parametrized, in a frame where 
$p = (E_{3}, p_{3} \cos\phi \sin\theta, p_{3} \sin\phi \sin\theta,
p_{3} \cos\theta)$, as
\begin{eqnarray}
\varepsilon(p_{3},0) &=& \frac{E_{3}}{m_{\omega}}
(\frac{p_{3}}{E_{3}}, \cos\phi \sin\theta, 
\sin\phi \sin\theta,\cos\theta) \,,\nonumber \\
\varepsilon(p_{3},\pm 1) &=& \frac{1}{\sqrt{2}}
(0, \mathrm{i} \sin\phi \mp \cos\theta\cos\phi,
-\mathrm{i} \cos\phi \mp \cos\theta\sin\phi,
\pm \sin\theta)\,.
\label{vectors}
\end{eqnarray}
It is easy to check that they fulfill the relation
$p^{\mu} \varepsilon_{\mu}(p,\lambda)$ = 0.

\subsection{\boldmath{$\omega$}-production as a diffractive excitation of the
\boldmath{$\omega p$}-Fock state}

The exclusive production of $\omega$-mesons in the fragmentation region 
of either proton can also be understood as 
a diffractive excitation of a two-body $\omega p$-Fock state
of the physical proton.
This is best formalized by a Fock-state decomposition of the
protons light-cone wave function in terms of meson-baryon
Fock states. A comprehensive treatment of meson-cloud effects
with applications to deep-inelastic scattering and
baryon form factors within this framework has been developed 
in \cite{HSS96,Dziembowski}, for a review and references see
\cite{Speth_Thomas}. 
For the problem at hand, we can write schematically
\begin{eqnarray}
\ket{p}_{phys} = \sqrt{Z} \Big( \ket{p}_{bare} + 
\int dz d^2 \vec{k}_\perp \, \Psi_{\omega p}(z,\vec{k}_\perp) 
\ket{p(1-z,-\vec{k_\perp}); \omega(z,\vec{k_\perp})} + \dots \Big) \, .
\end{eqnarray}
Here, the bare proton state represents, for example,
a three-quark core of the physical proton, 
$\Psi_{\omega p}$ is the light-cone wave function
of the $\omega p$-Fock-state. The $\omega$-meson 
in the two-body Fock-state
carries a fraction $z$ of light-cone plus-momentum 
of the physical proton
and transverse momentum $\vec{k}_\perp$; for simplicity helicity labels
are suppressed. 
The invariant mass of the virtual $\omega p$ system is then
given as
\begin{eqnarray}
M^2_{\omega p} = {\vec{k}_\perp^2 + m_\omega^2 \over z} +
{\vec{k}_\perp^2 + m_N^2 \over (1-z)} \, ,
\end{eqnarray}
and enters the radial part of the wave function in terms of the
$\omega NN$-form factor
\begin{eqnarray}
F_{\omega NN}(M^2_{\omega p}) = 
\exp \Big(-{M^2_{\omega p} - m_N^2 \over 2
\Lambda_{LC}^2}\Big) \, .
\label{eq:Formfactor}
\end{eqnarray}
The parameter $\Lambda^2_{LC}$ which controls the momentum 
distribution of $\omega$-mesons in the Fock-state is taken as
$\Lambda_{LC} = 1.1$ GeV \cite{HSS96}.

In accordance with the classic Good-Walker formalism
\cite{Good_Walker}, diffractive excitation of the $\omega p$-state now 
occurs because interactions of the bare proton and the 
two-body $\omega N$-state differ. 
We can write the $\omega p$ scattering state as:
\begin{eqnarray}
\ket{\omega p}_{scatt} = \Big( \hat{S}_{\omega p} - \hat{S}_p \Big) 
\ket{\omega p} \, ,
\label{eq:scatt_state}
\end{eqnarray}
where $\hat{S}_{\omega p}$ and $\hat{S}_p$ are the elastic 
scattering matrices for the $\omega p$ and $p$ interactions 
with the target.
Assuming, that the $S$-matrix of the two-body state factorizes,
$\hat{S}_{\omega p}= \hat{S}_{\omega} \, \hat{S}_{p}$, one can
show that Eq.(\ref{eq:scatt_state}) generates precisely the 
diagrams a), c), e) of Fig.\ref{fig:bremsstrahlung_diagrams}.
Diagrams b), d), f) can be obtained by an obvious symmetrization.
In the practical evaluation, these diagrams will give
similar expressions in momentum space as the ones obtained
in the reggeized field theory model (the ``standard approach''
discussed above), modulo the absence of Regge-factors and
the careful replacement of all $\omega NN$-form factors by
their light-cone counterparts given in Eq.(\ref{eq:Formfactor}).

Notice that this description of diffractive dissociation, which
treats the $\omega$-meson as a nonperturbative parton of the 
proton has a good physical motivation only in the fragmentation 
region of the proton(s). When the $\omega$-meson is produced
in the central rapidity domain, the reggeization of the
crossed channel exchanges must be taken into account.
For Reggeon exchanges however the light-cone wave function 
formalism described above is ill defined \cite{NSSS99}.
Therefore, for a description of midrapidity $\omega$ production,
one would have to add the reggeized $\omega$ exchange.
We do not do this here, as the final result would not 
differ much from the reggeized field theory diagrams (the ``standard
approach''). At rapidities close to the proton fragmentation
region the difference between the ``standard approach'' and
the light-cone wave function treatment can serve
as an indicator for the model dependence of our predictions 
for this particular soft process.

Finally let us note, that at the high energies of interest 
the deviation from factorization
\begin{eqnarray}
\delta \hat{S} = \hat{S}_{\omega p} -\hat{S}_{\omega} \, \hat{S}_{p} \, ,
\end{eqnarray}
is quantified by the shadowing or absorption correction to 
which we now turn.

\subsection{Absorption effects}
\begin{figure}[!h]
\includegraphics[width = 4.5cm]{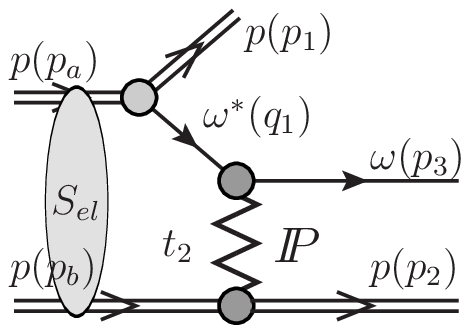}
\includegraphics[width = 4.5cm]{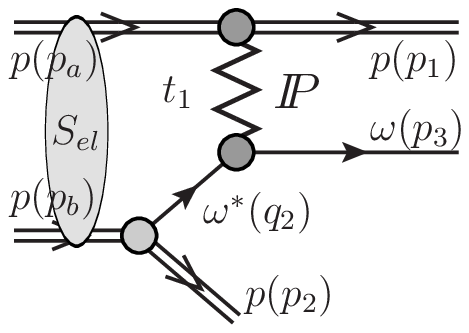}
  \caption{\label{fig:bremsstrahlung_absorption_diagrams}
  \small The absorption effects included in the present paper 
for the $\omega$ bremsstrahlung.
}
\end{figure}

The absorption effect for the hadronic bremsstrahlung contributions requires a short
comment. Since in practice for the pomeron exchanges in diagrams
a) - d) we use phenomenological interactions which effectively describe 
the total and elastic data an additional use of absorption would be
a double counting.
This is not the case for diagrams e) and f) where
the interaction is between $\omega$-meson and proton. Consequently in the latter
case we include absorption effect in full analogy to that described
in section about photoproduction. This is illustrated in 
Fig.\ref{fig:bremsstrahlung_absorption_diagrams}.

\section{Results}

In the present section we present differential distributions
for three different energies: $W = 200$ GeV (RHIC),
$W = 1960$ GeV (Tevatron) and $W = 14$ TeV (LHC). 
This includes rapidity and transverse momentum of $\omega$ meson 
distributions as well as
azimuthal correlations between outgoing protons.

\begin{figure}[!h]
\includegraphics[width = 0.32\textwidth]{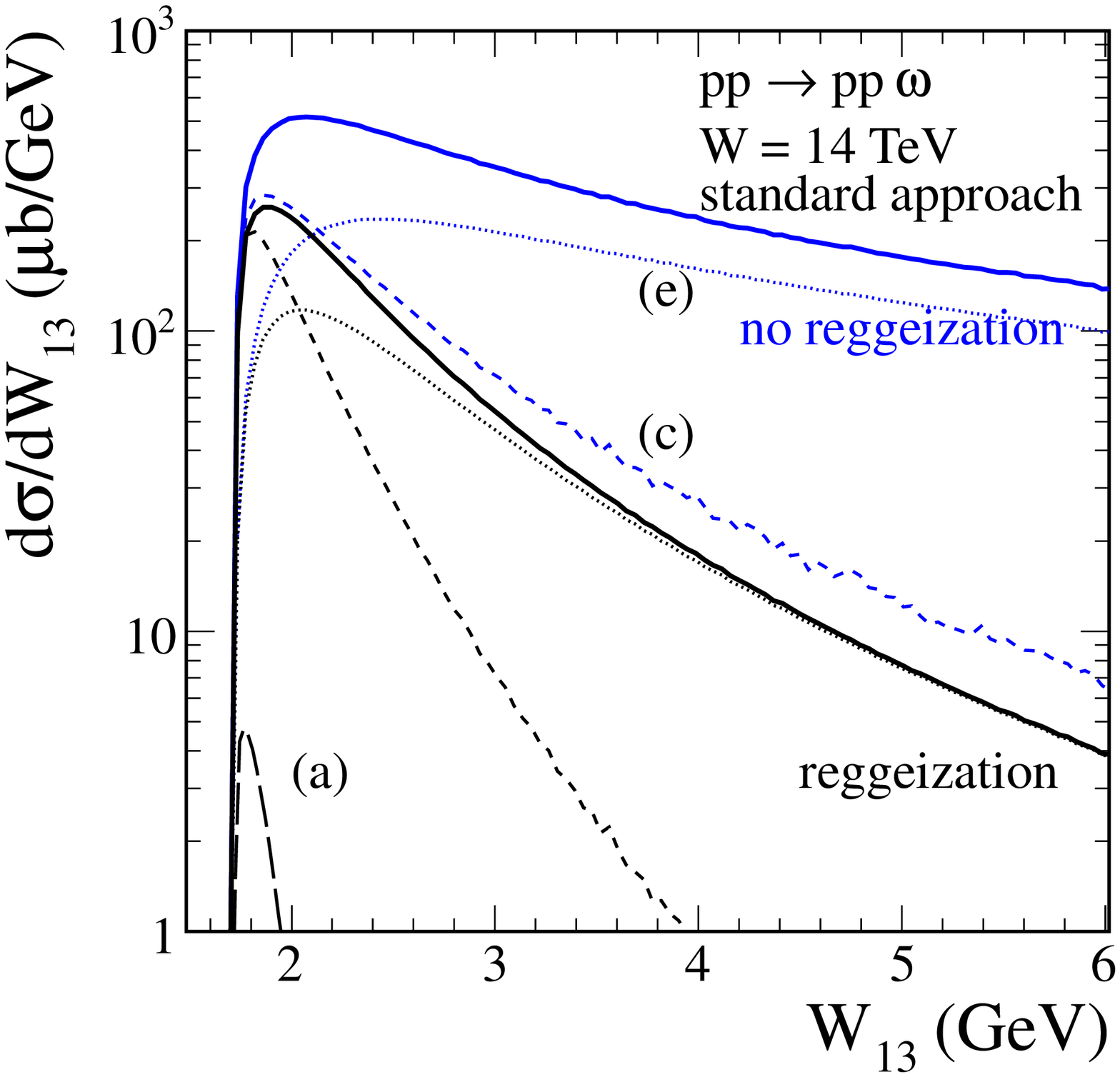}
\includegraphics[width = 0.32\textwidth]{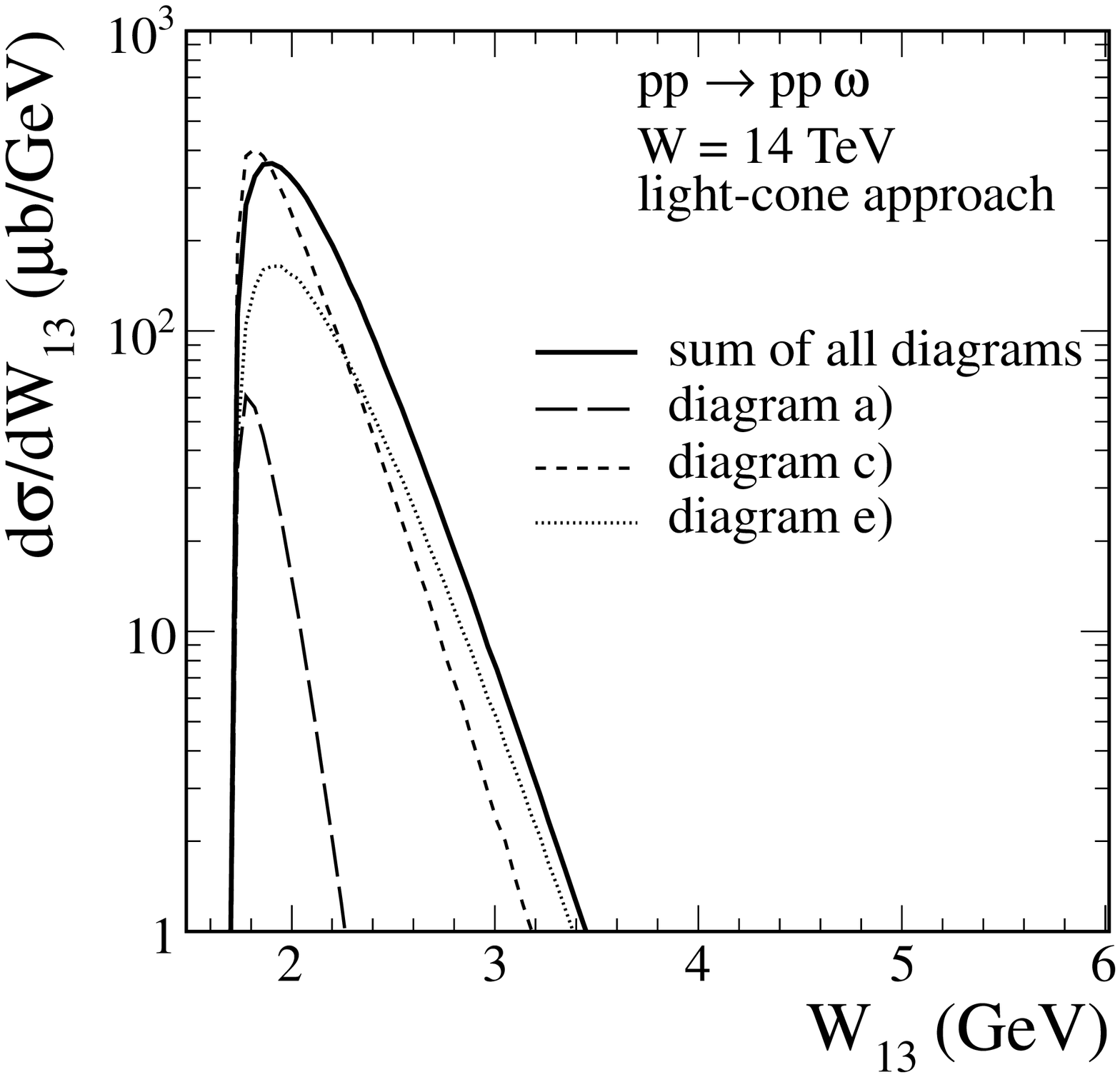}
  \caption{\label{fig:dsig_dw13}
  \small
Differential cross sections $d\sigma/dW_{13}$
for the $pp \to pp \omega$ reaction at $W$ = 14 TeV
for the hadronic bremsstrahlung mechanisms.
The left panel is for results with Mandelstam variable dependents
$\omega NN$ form factors and with reggeization included
while the light-cone approach correspond to the right panel.
The thick solid line presents the result for the coherent sum
of all amplitudes shown in Fig.\ref{fig:bremsstrahlung_diagrams}.
}
\end{figure}
\begin{figure}[!h]
\includegraphics[width = 0.32\textwidth]{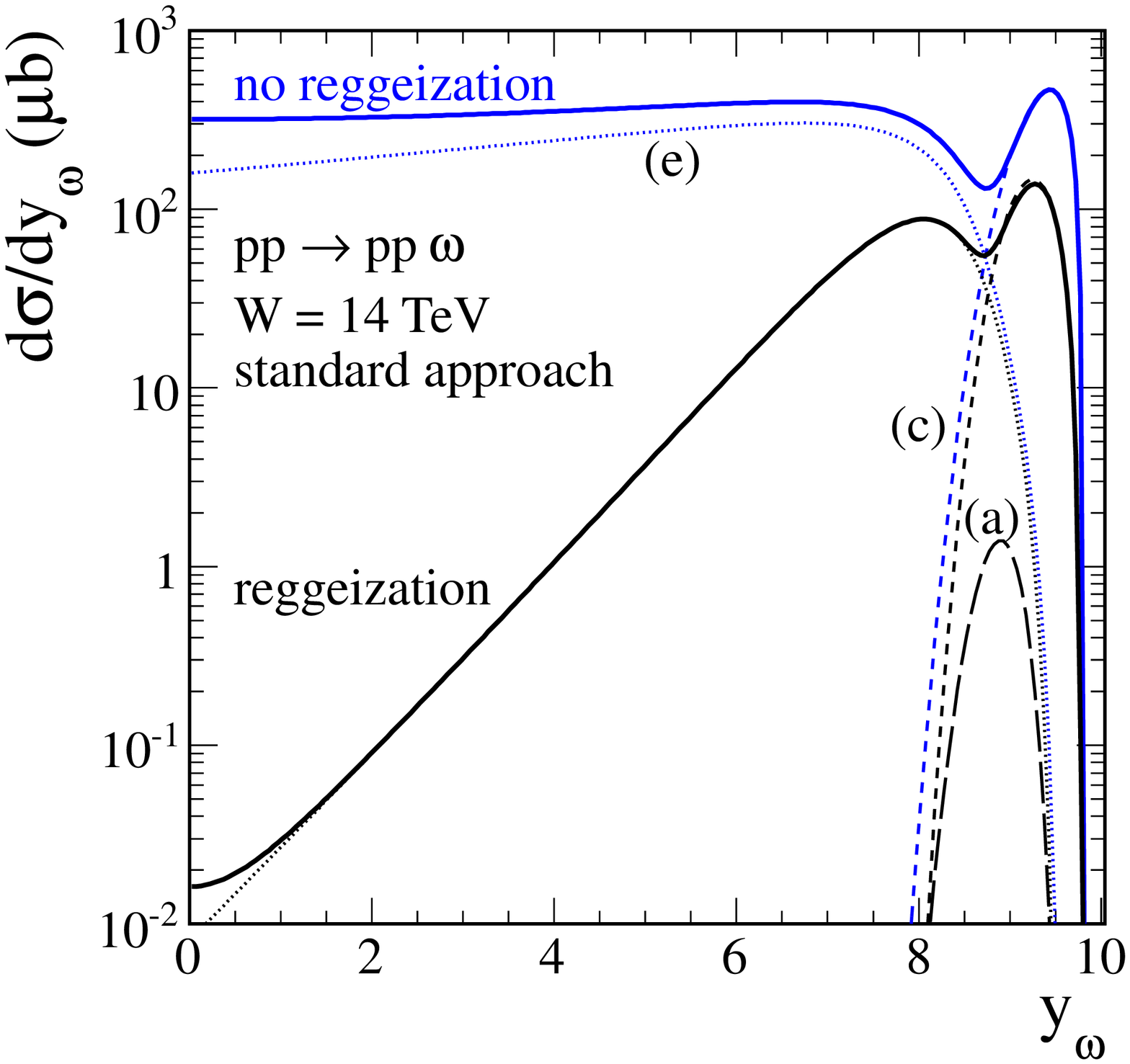}
\includegraphics[width = 0.32\textwidth]{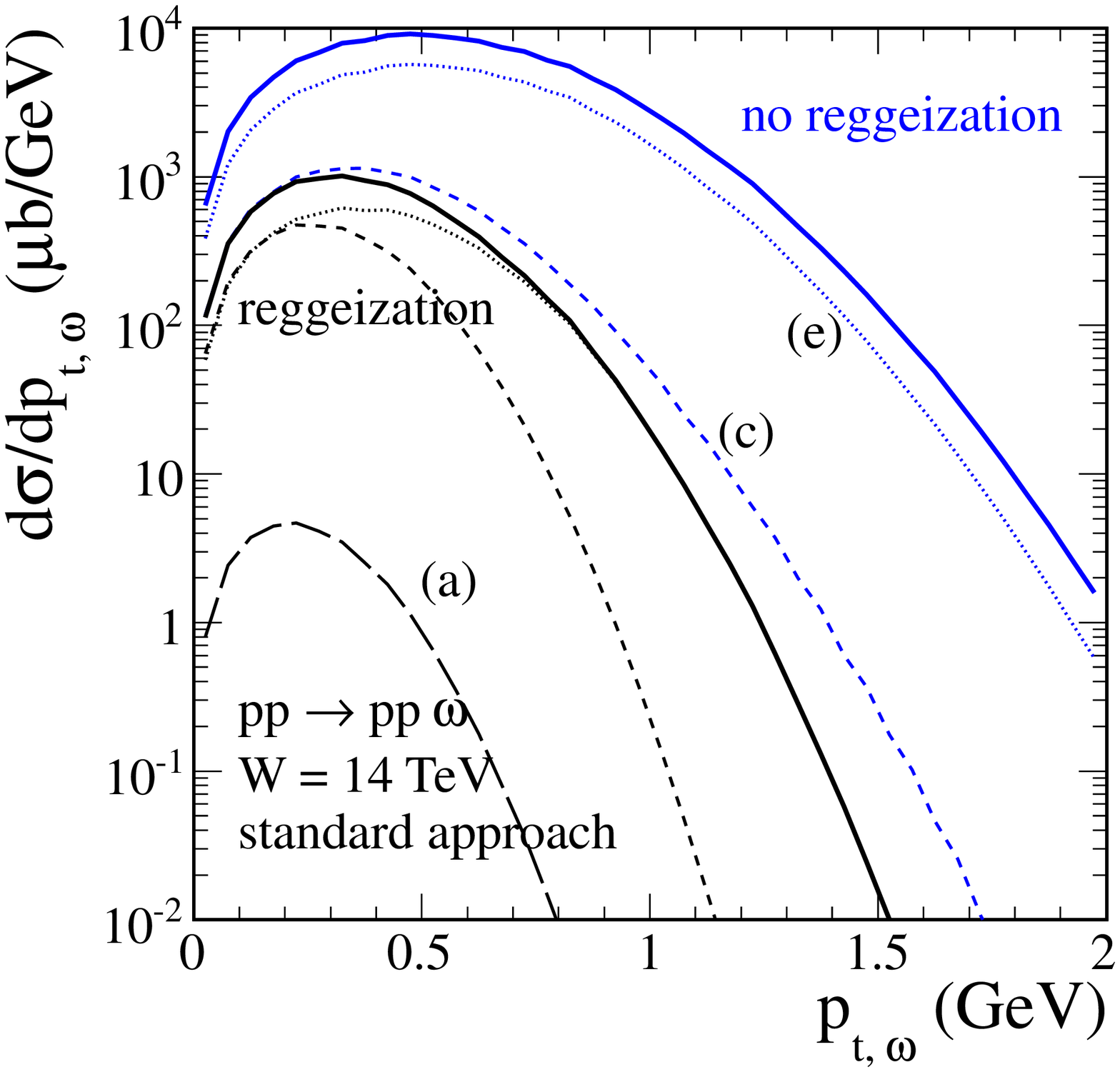}
\includegraphics[width = 0.32\textwidth]{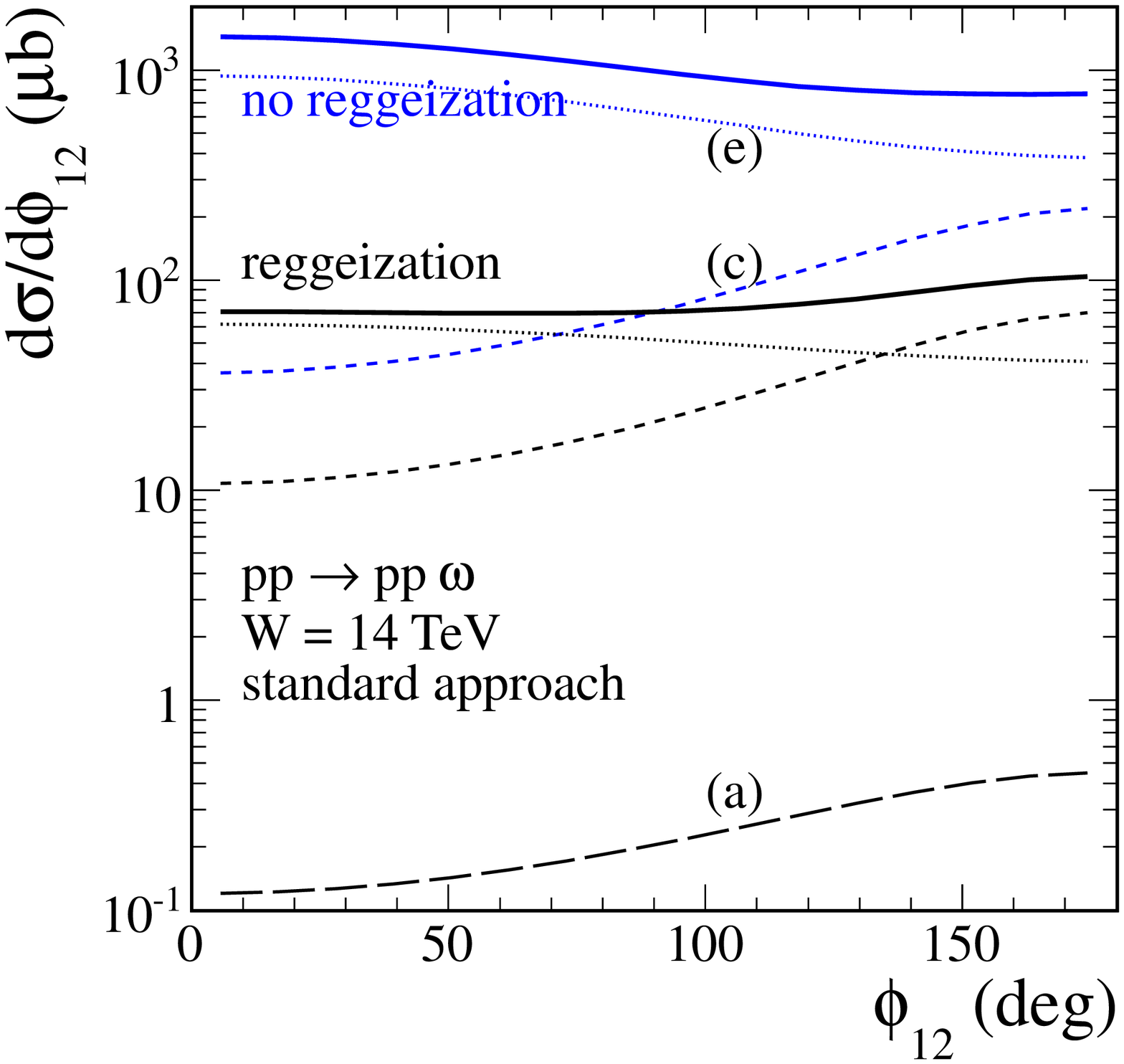}\\
\includegraphics[width = 0.32\textwidth]{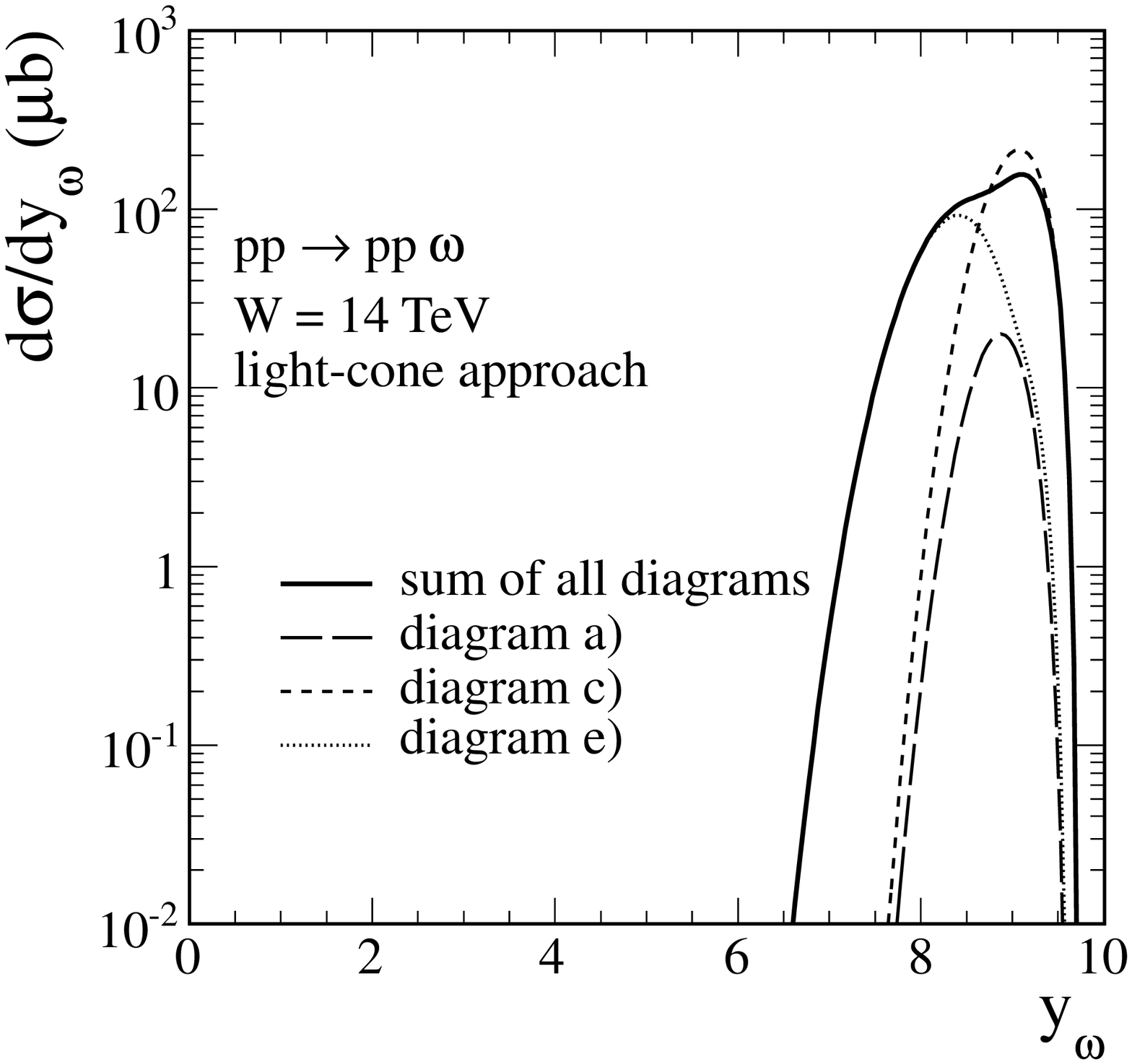}
\includegraphics[width = 0.32\textwidth]{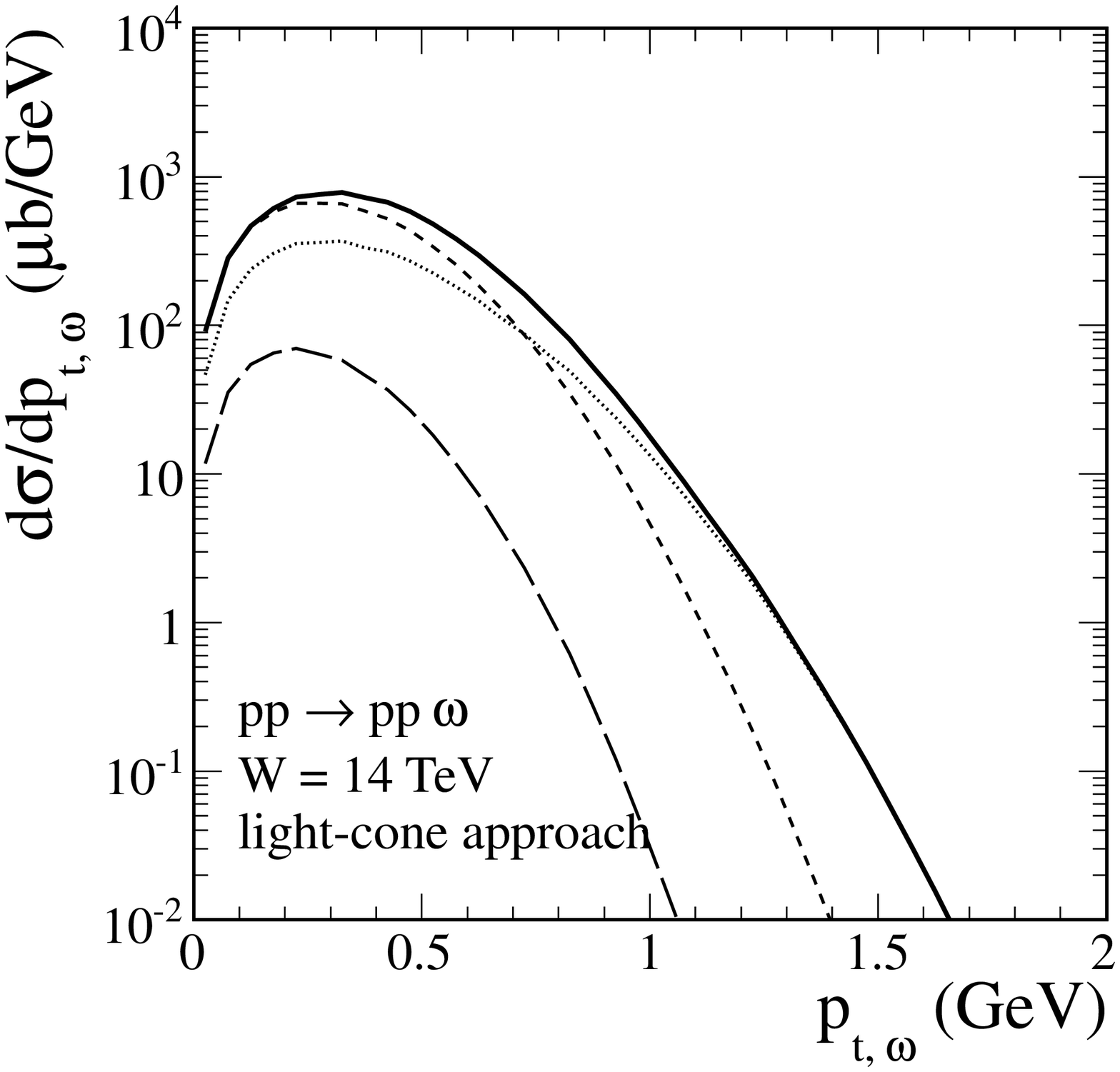}
\includegraphics[width = 0.32\textwidth]{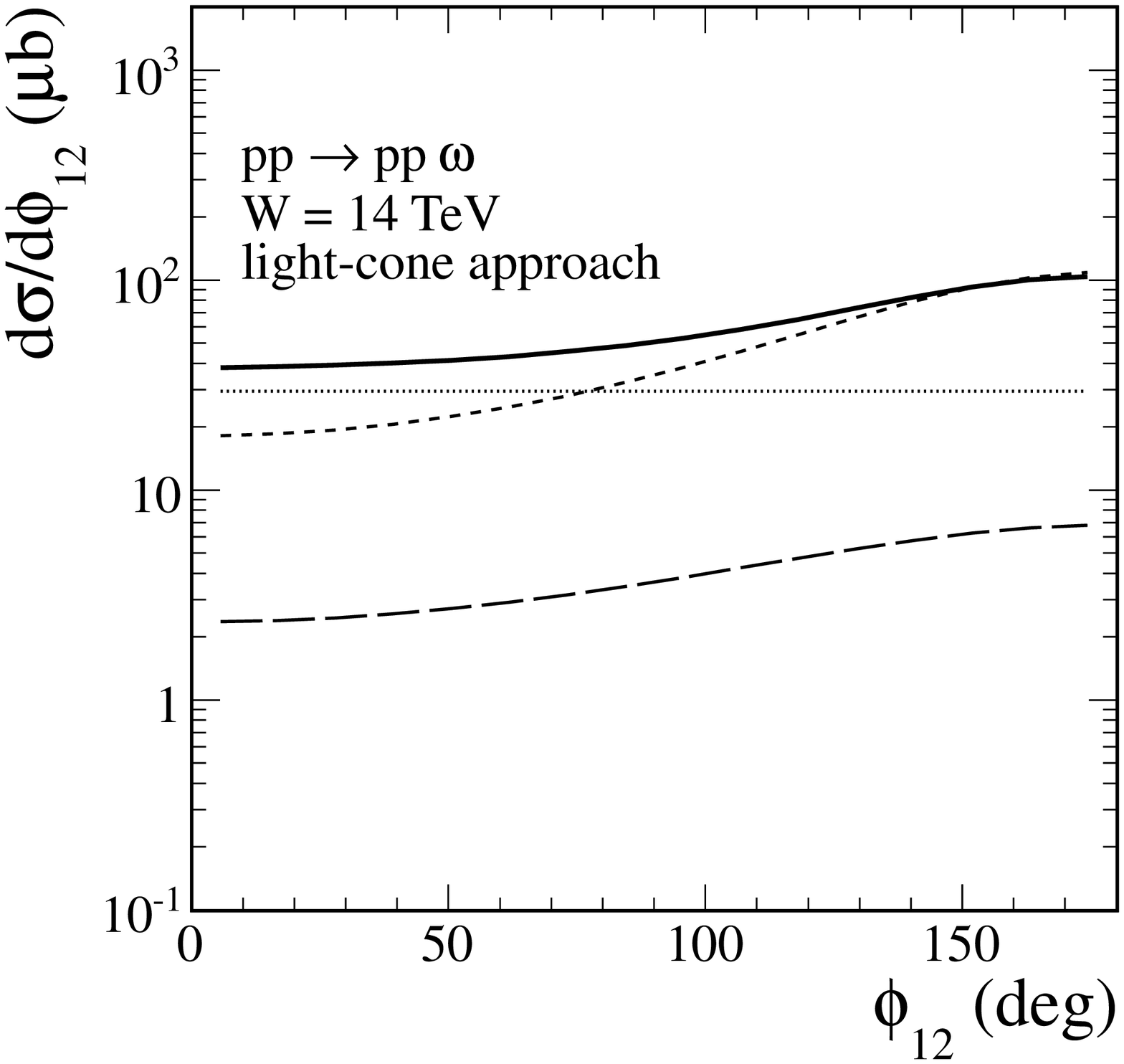}
  \caption{\label{fig:dsig_deco}
  \small
Differential cross sections
for the $pp \to pp \omega$ reaction at $W = 14$ TeV
for the hadronic bremsstrahlung mechanisms.
The upper panels are for results with Mandelstam variable dependents
$\omega NN$ form factors and with reggeization included
while the light-cone
form factors correspond to the bottom panels.
The thick solid line presents the cross sections for the coherent sum
of all amplitudes shown in Fig.\ref{fig:bremsstrahlung_diagrams}.
}
\end{figure}

In Fig.\ref{fig:dsig_dw13} we present 
differential cross sections $d\sigma/dW_{13}$
for the $pp \to pp \omega$ reaction at $W = 14$ TeV.
We show results with Mandelstam variable dependent
form factors (left panel),
which we will call standard in the following,
and with light-cone form factors (right panel).
In the left panel we show results for the standard spin-1/2 propagators
in diagrams a) and c) as well as with reggezaition \cite{Storrow}. 
The long dashed, dashed and dotted lines correspond to
contributions from diagrams a), c) and e), respectively.
The thick solid line presents the coherent sum
of all amplitudes.
The light-cone form factors
lead to much steeper dependence of the cross section on $W_{13}$ ($W_{23}$)
than the standard form factors.
The reggezaition leads to an extra damping of the large $W_{13}$ ($W_{23}$)
cross section.

In Fig.\ref{fig:dsig_deco} we present the role of the form factors and reggezaition
for differential distributions in the $\omega$ meson rapidity and transverse momentum
as well as for azimuthal angle correlation between outgoing protons.
The distribution in rapidity is closely related to that for $W_{13}$ ($W_{23}$).
As seen from the middle panels the reggezaition makes the distribution steeper
in the $\omega$ meson transverse momentum.

\begin{figure}[!h]
\includegraphics[width = 0.32\textwidth]{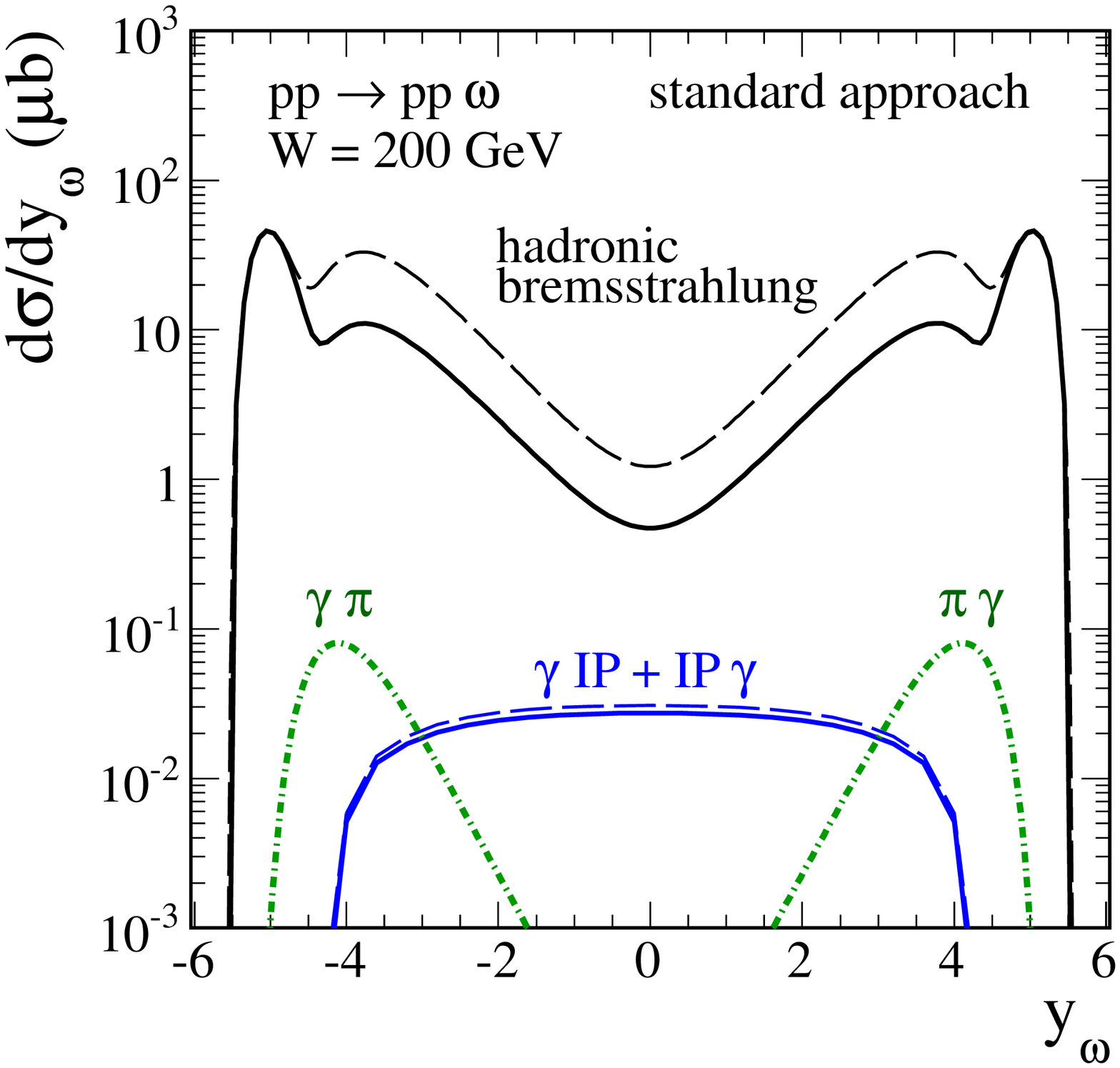}
\includegraphics[width = 0.32\textwidth]{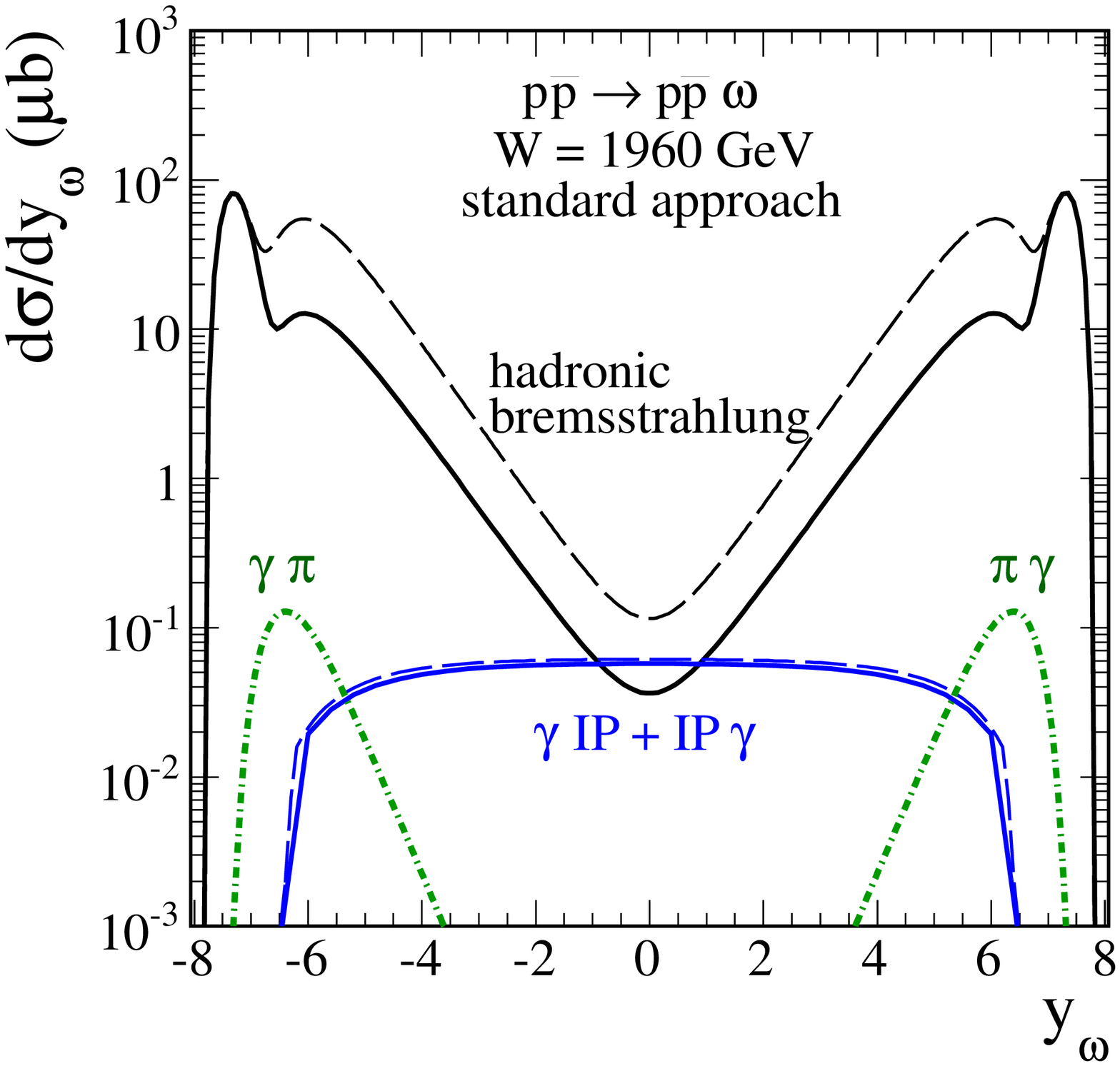}
\includegraphics[width = 0.32\textwidth]{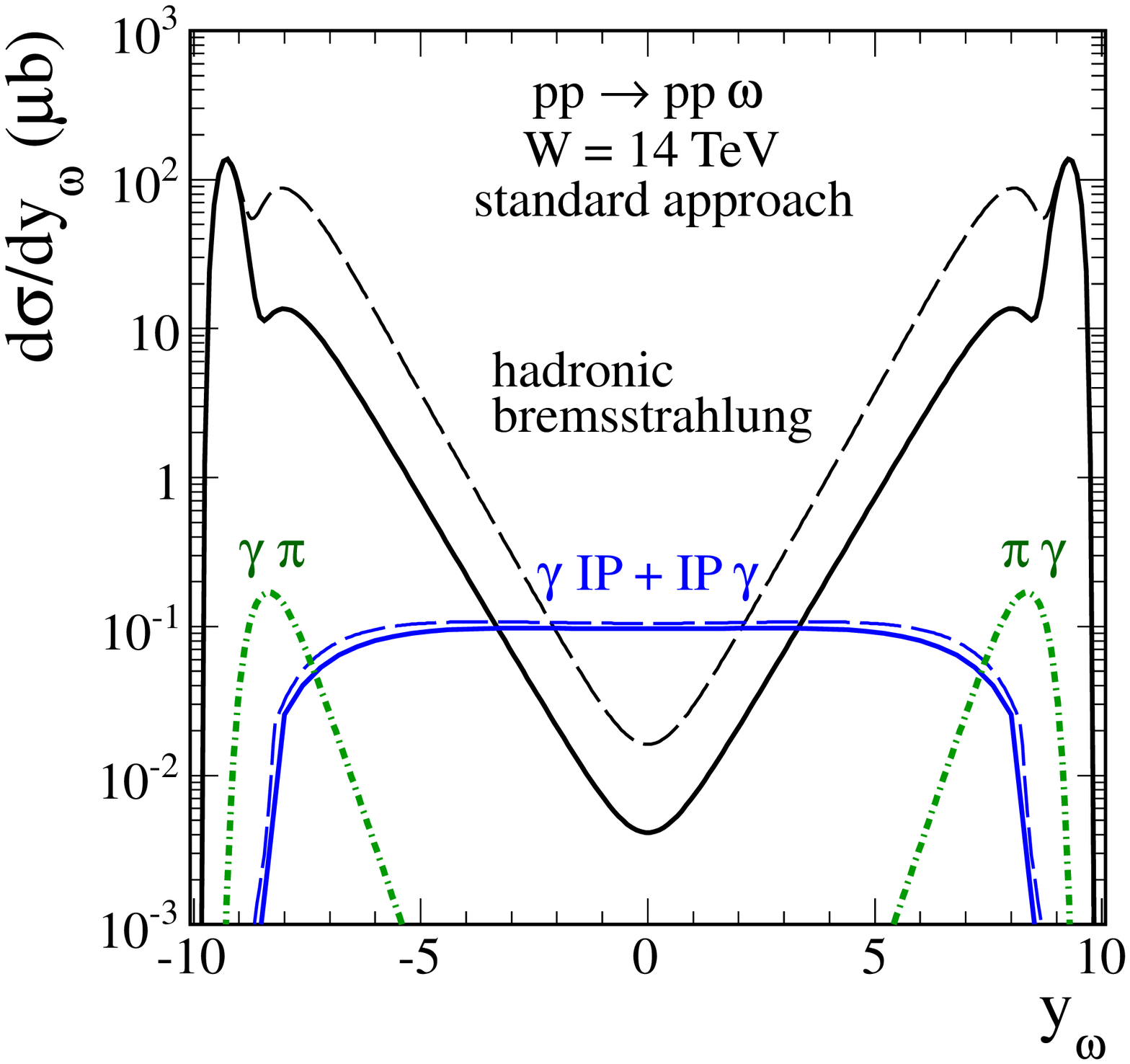}\\
\includegraphics[width = 0.32\textwidth]{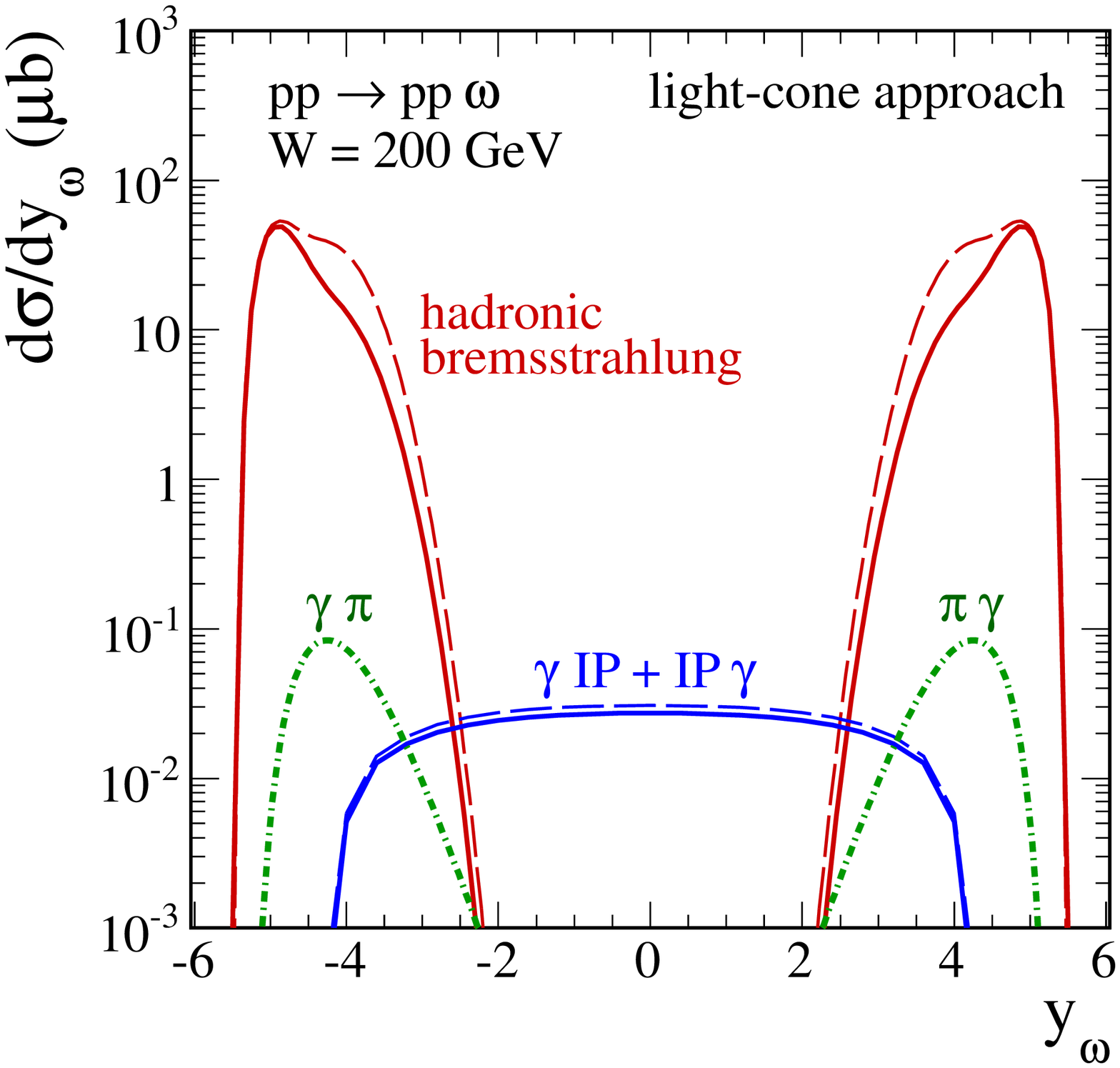}
\includegraphics[width = 0.32\textwidth]{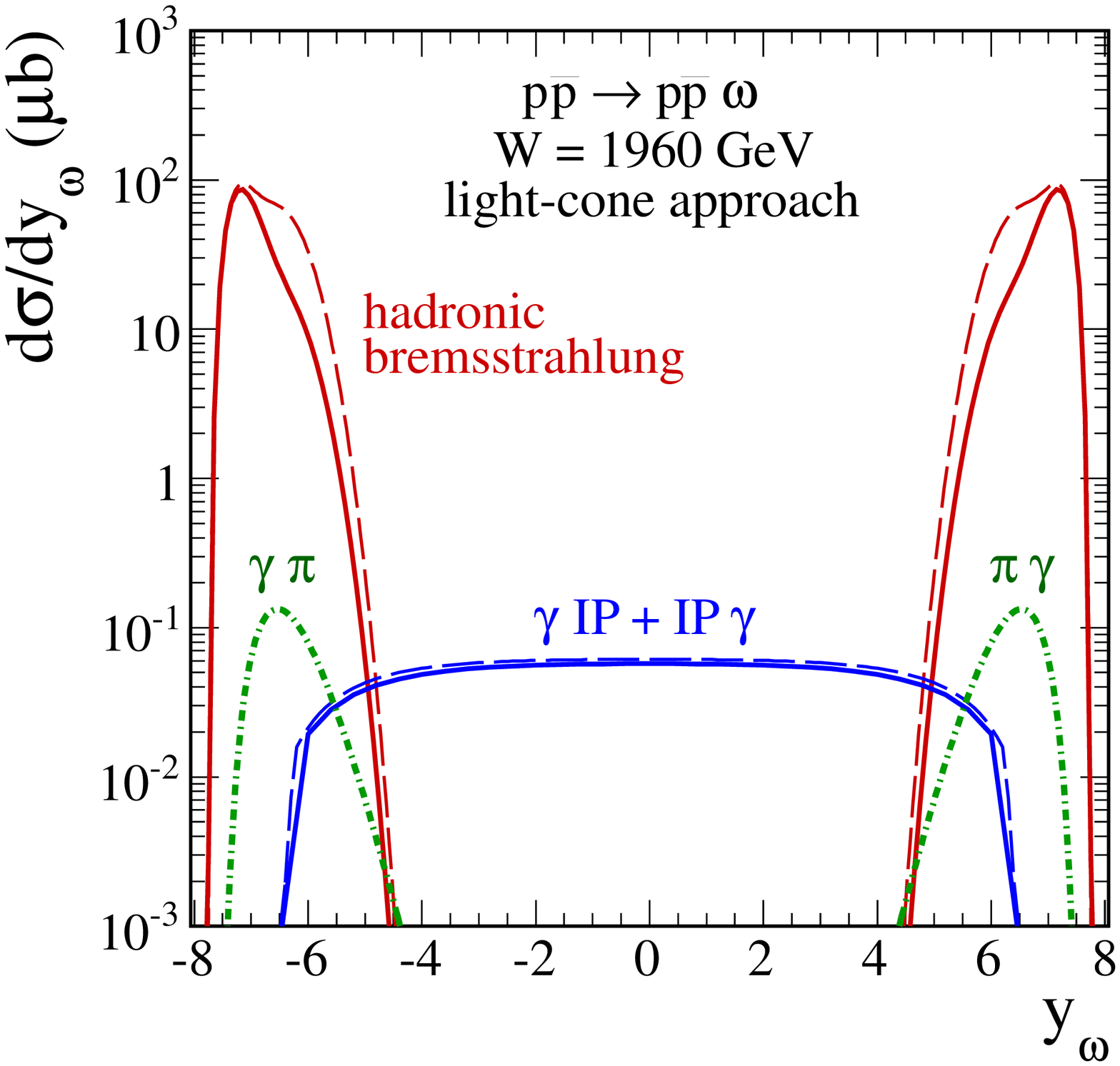}
\includegraphics[width = 0.32\textwidth]{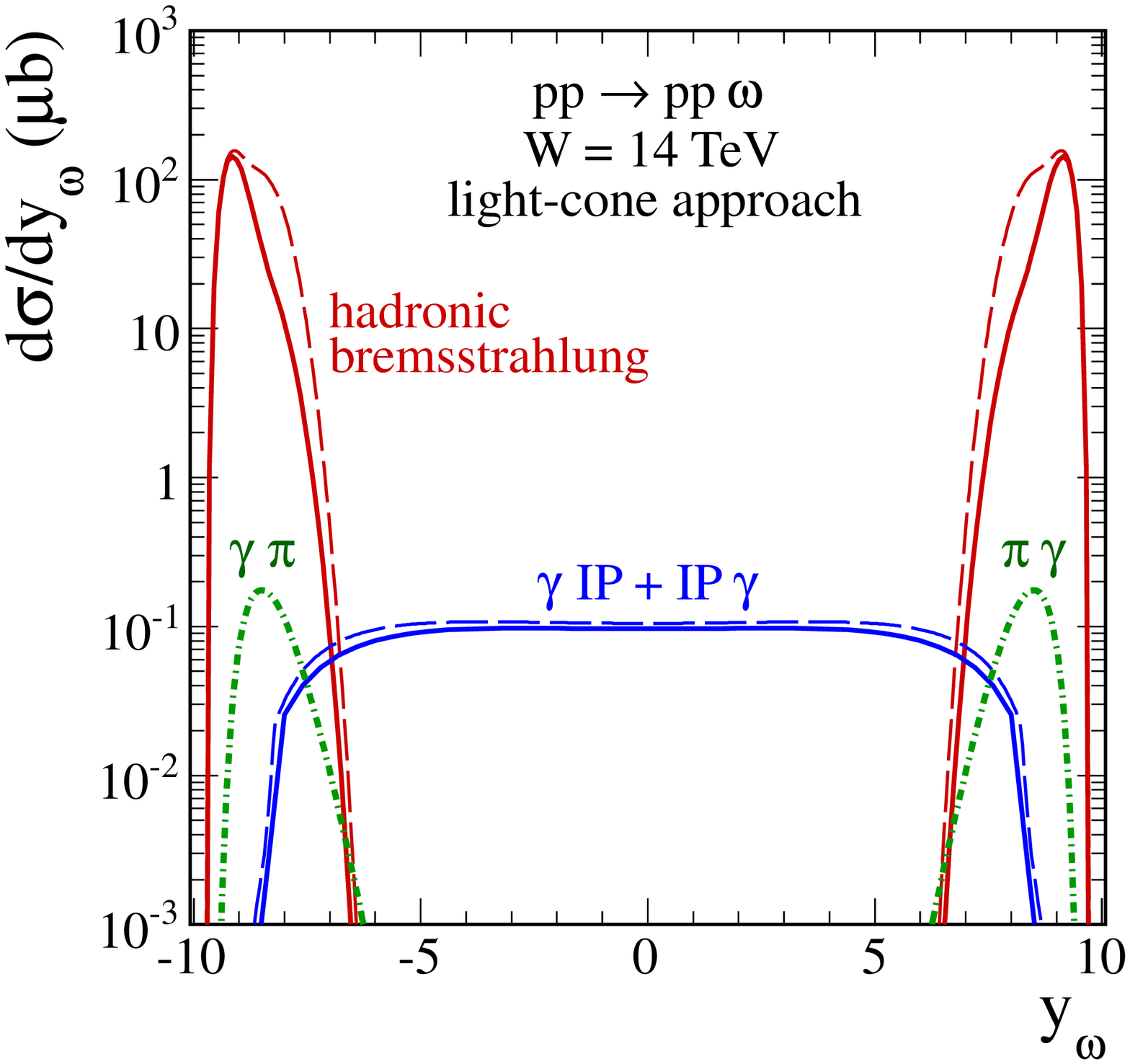}
  \caption{\label{fig:dsig_dy}
  \small
Differential cross sections $d\sigma/dy_{\omega}$
for the $pp(\bar{p}) \to pp(\bar{p}) \omega$ reaction
at $W$ = 200, 1960, 14000 GeV in the full rapidity range.
The upper panels are for results with Mandelstam variable dependent
$\omega NN$ form factors and with reggeization included
while the light-cone
form factors correspond to the bottom panels.
In the latter case the reggeon exchanges are evidently not included.
The black lines present the contribution of the
hadronic bremsstrahlung diagrams.
The blue lines correspond to the QCD $\gamma I\!\!P$ and $I\!\!P \gamma$ mechanism.
The green dash-dotted lines present the contribution of diagrams 
for the $\gamma \pi^0$ (left peak) and $\pi^0 \gamma$ (right peak) exchanges.
The dashed lines in the figures present the contributions without absorption,
while the thick solid lines include the absorption.
}
\end{figure}

In Fig.\ref{fig:dsig_dy} we present rapidity distribution of the $\omega$ meson 
in the two approaches for different energies.
In the first approach we use the standard $\omega N N$ form 
factors (upper panels) 
and in the second approach we use the light-cone
form factors (bottom panels) for the omega-nucleon-nucleon coupling.
The distributions for the standard form factors
extend more towards midrapidities.
We show the $\gamma I\!\!P$ ($I\!\!P \gamma$),
$\gamma \pi^0$ ($\pi^0 \gamma $)
as well as diffractive bremsstrahlung mechanisms.
At ``low'' energy (RHIC) the discussed hadronic bremsstrahlung mechanisms dominate
over the $\gamma I\!\!P$ and $I\!\!P \gamma$ ones. The cross section for 
the hadronic bremsstrahlung contribution is two-orders of magnitude bigger
than that for the ($\gamma I\!\!P,
I\!\!P \gamma$) contribution. The latter mechanism is known to be the
dominant one for $J/\Psi$ and $\Upsilon$ meson production \cite{SS07,RSS08}.
A recent analysis at the Tevatron seems to confirm this claim
\cite{Tevatron_JPsi}.
Increasing the center-of-mass energy the hadronic bremsstrahlung components move
to large rapidities. 
The $\gamma \pi^0$ (left peak)
and the $\pi^0 \gamma$ (right peak) components are separated.
The separation in rapidity means also lack of interference effects
which is very different compared to the $\gamma I\!\!P$ ($I\!\!P \gamma$)
mechanism.
\footnote{The interference beetwen the two mechanisms
$\gamma I\!\!P$ and $I\!\!P \gamma$ 
is proportional to $e_{1} e_{2} (\vec{p_{1}} \cdot \vec{p_{2}})$
and introduces a charge asymmetry as well as an angular 
correlations between the outgoing protons.}

At LHC energy at midrapidities
the photoproduction mechanisms with $I\!\!P$ exchange
dominate over the hadronic bremsstrahlung ones.
We predict a narrow plateau around $y \approx$ 0 and a significant
increase when going to large $|y|$.
Experimental observation of the increase would confirm 
the bremsstrahlung mechanisms discussed here.
Only at the highest LHC energy the region of very small
rapidities is free of the hadronic bremsstrahlung contributions.

The difference between the results with standard and light-cone form factors
illustrates theoretical uncertainties. 
While the hadronic bremsstrahlung contributions are subjected to rather large theoretical
uncertainties (see discussion above),
the $\gamma I\!\!P$ ($I\!\!P \gamma$) contributions are fairly precisely estimated.
Deviations from the pQCD contribution at midrapidities may
be caused by either the difficult to predict hadronic bremsstrahlung contributions
or by the very interesting pomeron-odderon contributions.
The rise of the cross section with increasing $|y|$ 
would be a clear signal of the hadronic bremsstrahlung contributions,
while a sizeable deviation of the cross section normalization a potential signal of 
the odderon exchange. 

\begin{figure}[!h]  
\includegraphics[width = 0.32\textwidth]{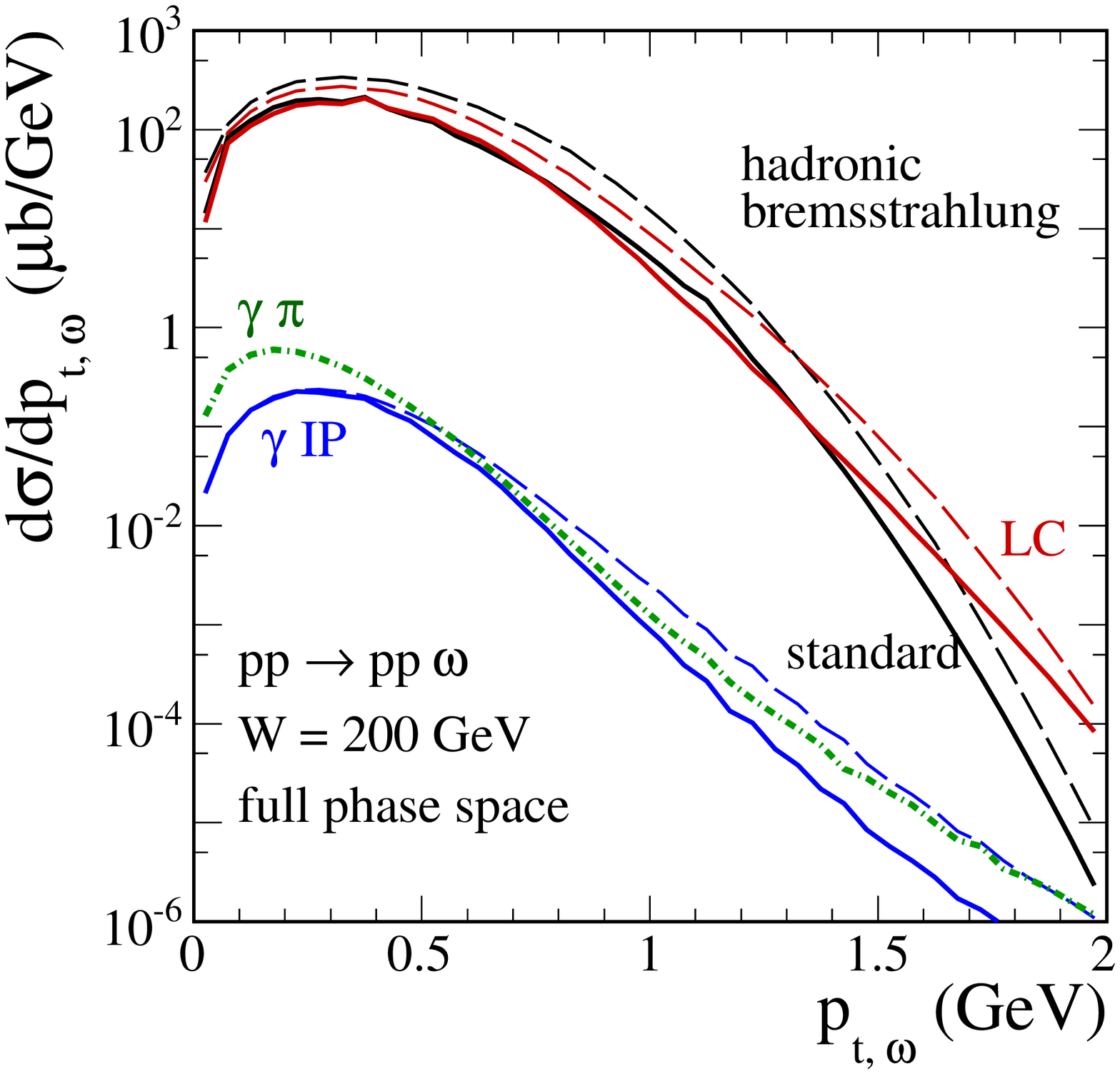}
\includegraphics[width = 0.32\textwidth]{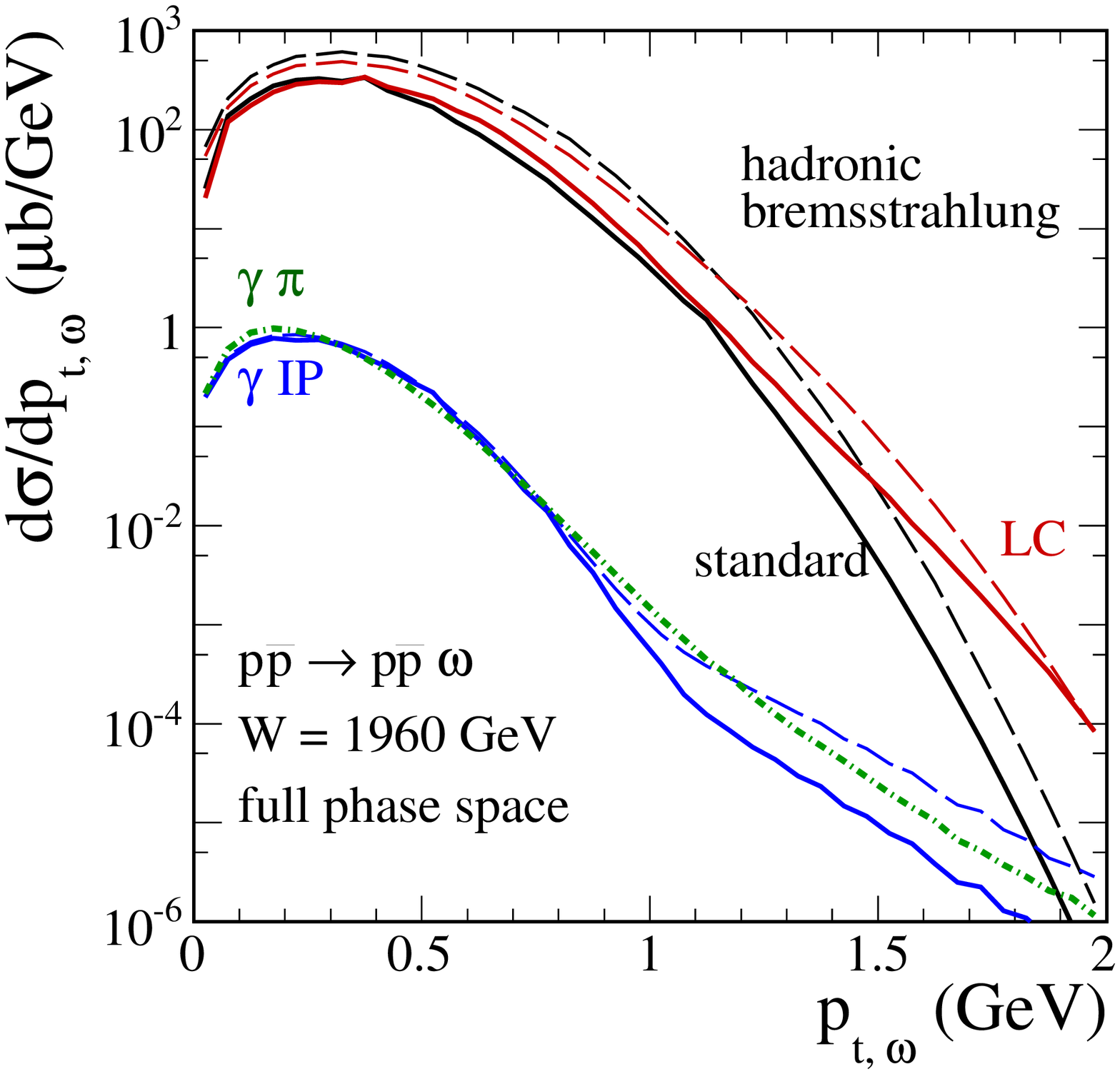}
\includegraphics[width = 0.32\textwidth]{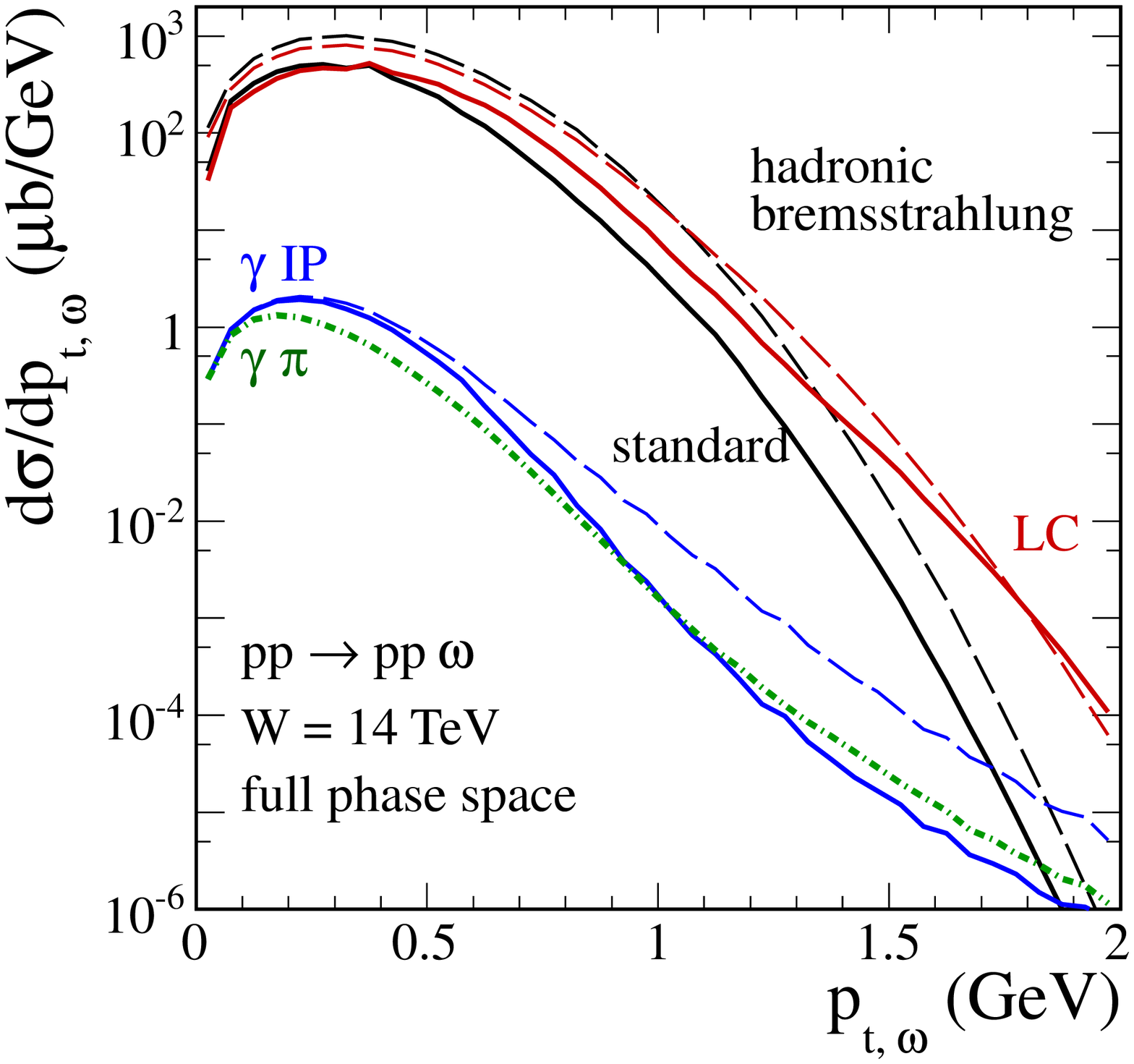}
  \caption{\label{fig:dsig_dpt}
  \small
Differential cross section $d\sigma/dp_{t}$
for the $pp(\bar{p}) \to pp(\bar{p}) \omega$ reaction
at $W$ = 200, 1960, 14000 GeV in the full rapidity range.
Here the reggeized propagators of omega and nucleons are used.
The dashed lines present the contribution without absorption,
while the thick solid lines include the absorption.
}
\end{figure}

In Fig.\ref{fig:dsig_dpt} we show the distribution in the $\omega$ meson
transverse momentum. In this case the integration is done over full
range of meson rapidities.
The thin lines are for the Born level calculations while the thick lines
include effect of absorption. The hadronic bremsstrahlung contributions calculated
with the light-cone form factors are steeper than those for the standard form factors.
The distribution of the photon-pomeron contribution for the $p \bar p$ scattering 
is somewhat different than that for the $p p$ scattering. This is caused by
different signs of the interference terms (different combination of electric charges).
The distribution of the $\gamma \pi^0$ ($\pi^0 \gamma$) contribution
(green dash-dotted line) is very similar to that of the $\gamma I\!\!P$ ($I\!\!P \gamma$)
contribution (blue lines).

\begin{figure}[!h]
\includegraphics[width = 0.32\textwidth]{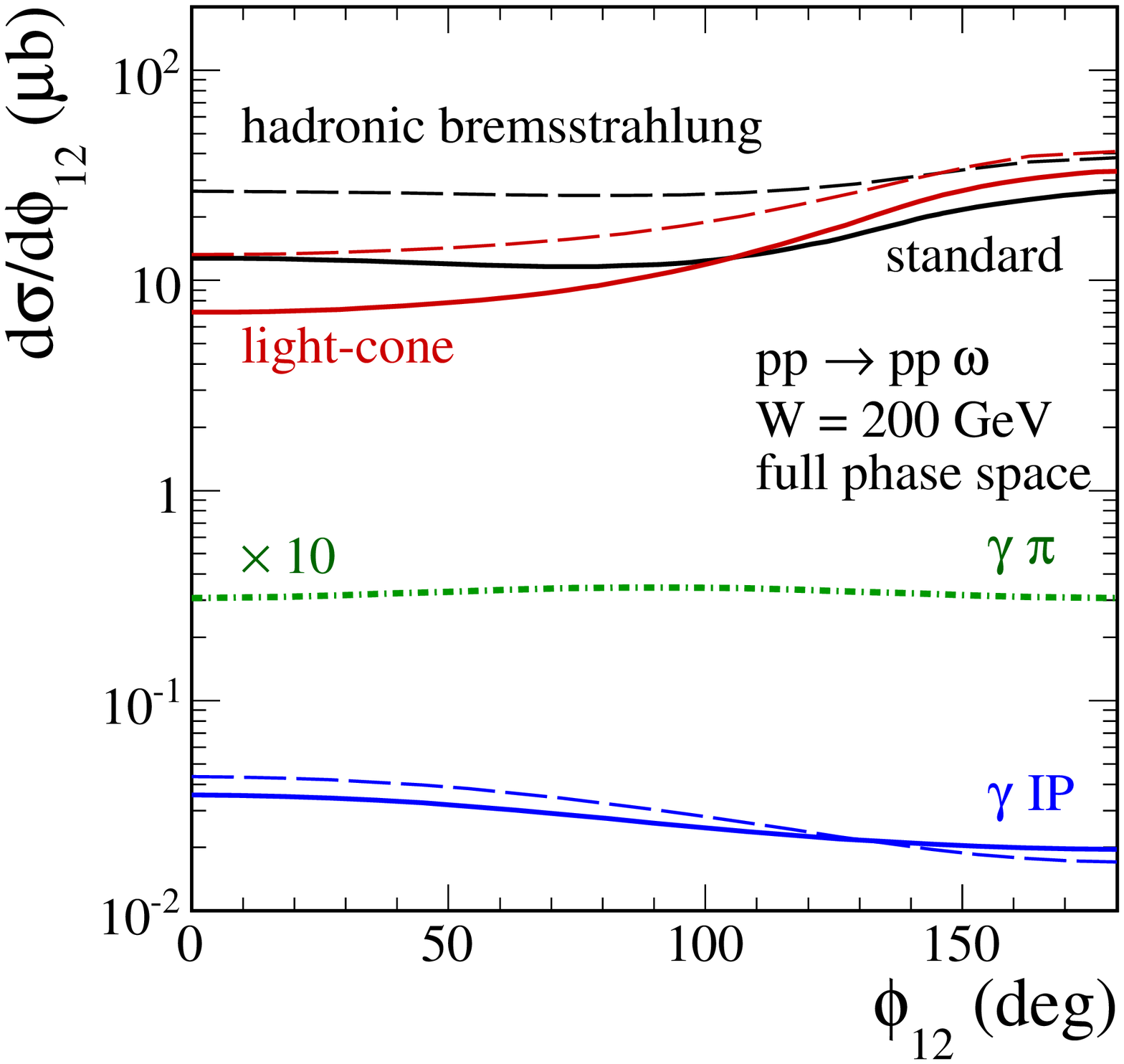}
\includegraphics[width = 0.32\textwidth]{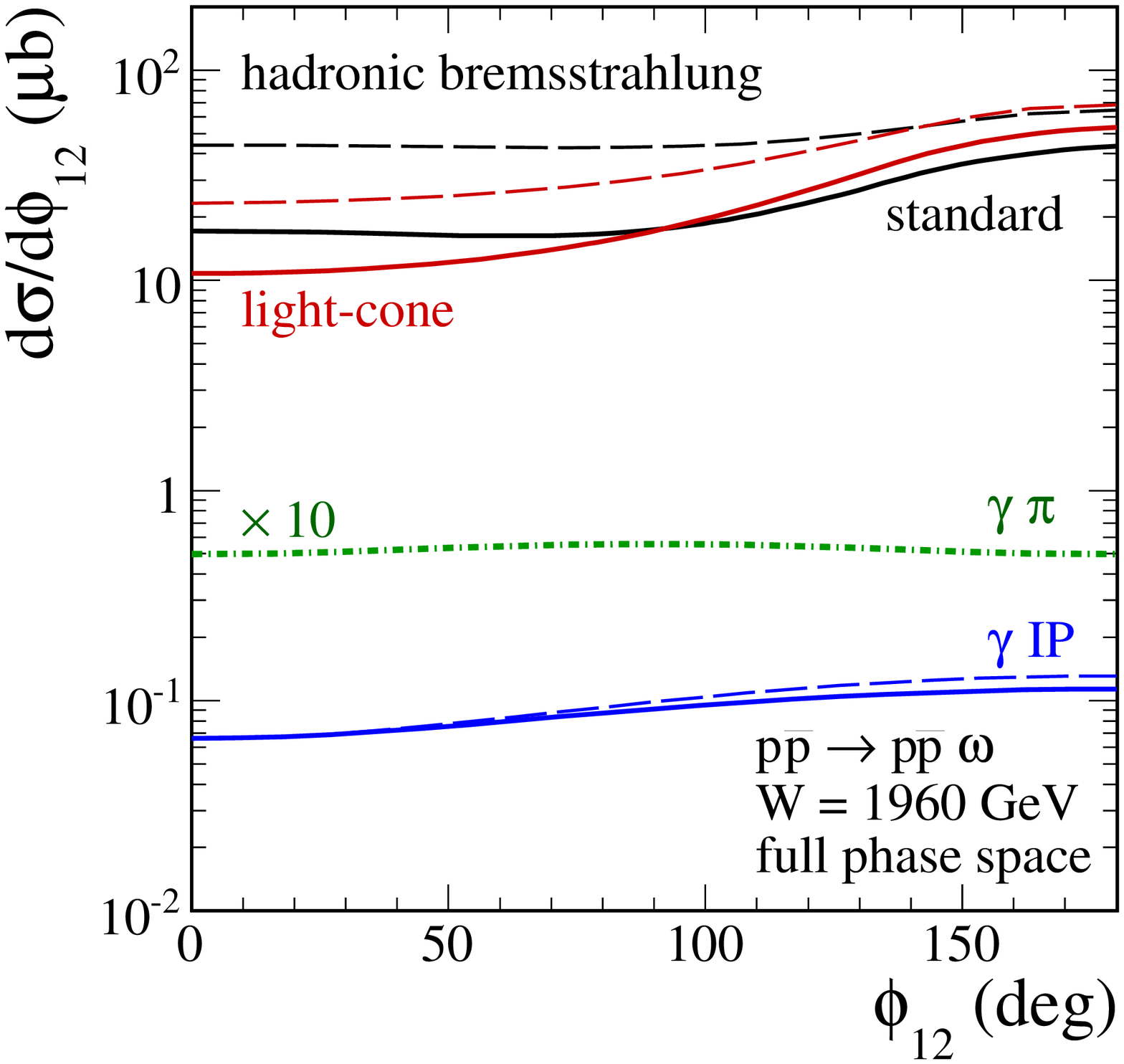}
\includegraphics[width = 0.32\textwidth]{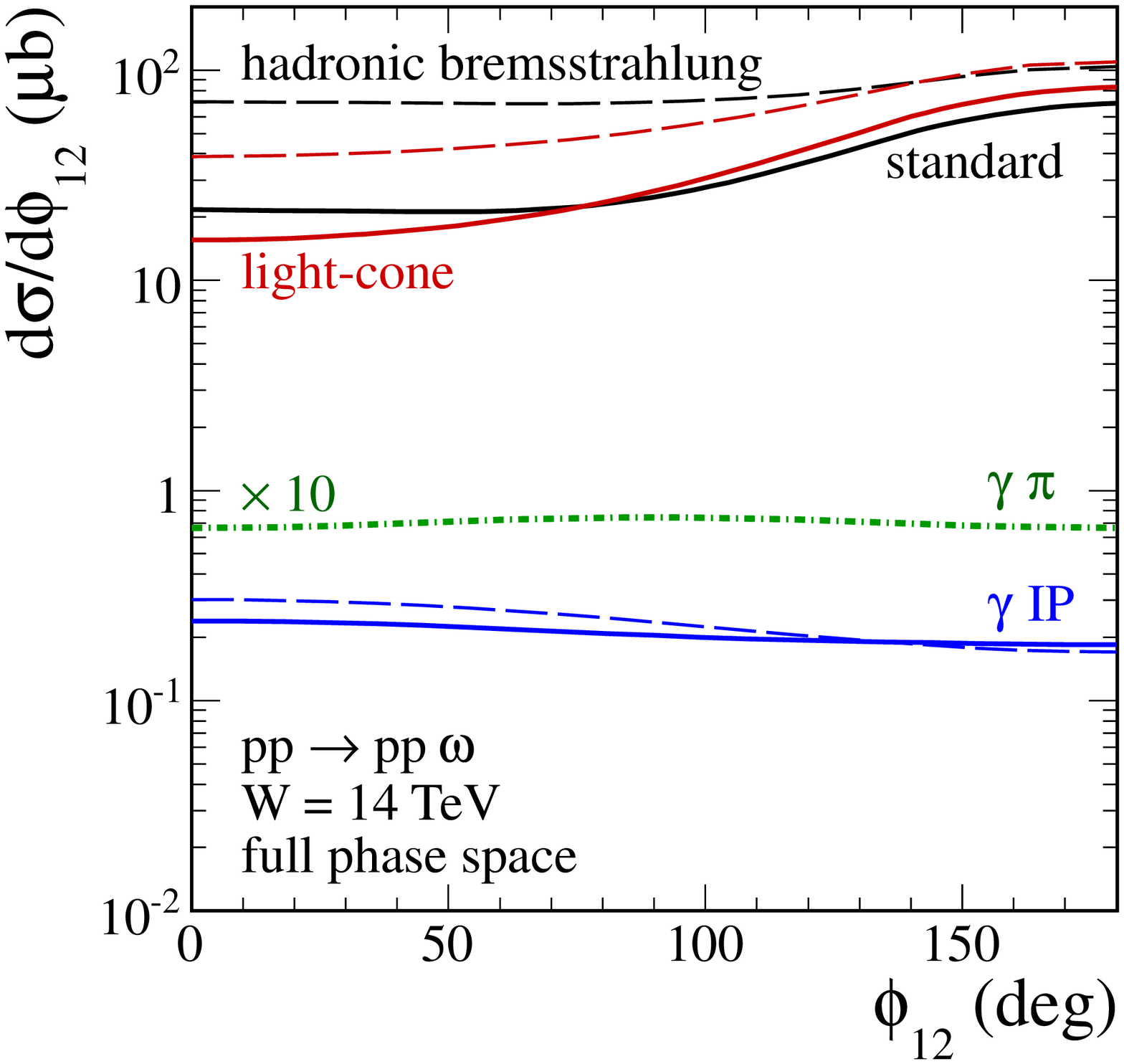}
  \caption{\label{fig:dsig_dphi}
  \small
Differential cross section $d\sigma/d\phi_{12}$
for the $pp(\bar{p}) \to pp(\bar{p}) \omega$ reaction
at $W$ = 200, 1960, 14000 GeV in the full rapidity range.
Here the reggeized propagators of omega and nucleons are used.
The dashed lines present the contribution without absorption,
while the thick solid lines include the absorption.
}
\end{figure}

Whether the $\gamma \pi^0$ mechanism can be identified requires further 
studies.
What are other specific features of this mechanism ?

In Fig.\ref{fig:dsig_dphi} we show distribution in relative
azimuthal angle between outgoing protons.
For the $\gamma \pi^0$ mechanism the maximum occurs at
$\phi_{12} \approx$ $\pi$/2 which is dictated by a specific tensorial coupling
$\gamma \pi^0 \to \omega$.
The azimuthal distribution for the $\gamma \pi^0$ mechanism is
very different than for the hadronic bremsstrahlung contributions
which peak at $\phi_{12}$ = $\pi$, especially for the light-cone form factors.
In principle, the azimuthal angle correlations could be used
therefore to separate the different mechanisms.
One can clearly see that the absorption effects lead
to extra decorrelation in azimuth compared to the Born-level result.
In Fig.\ref{fig:dsig_dphi} we show rapidity-integrated results.
In general the azimuthal angle correlations are rapidity dependent.
Quite different distributions for the $\gamma I\!\!P$ ($I\!\!P \gamma$) contribution
have been predicted for the Tevatron and RHIC or LHC.
The correlation function is for this mechanism
caused totally be the interference of the
$\gamma I\!\!P$ and $I\!\!P \gamma$ contributions
(see \cite{SS07}).

\begin{figure}[!h]  
\includegraphics[width = 0.32\textwidth]{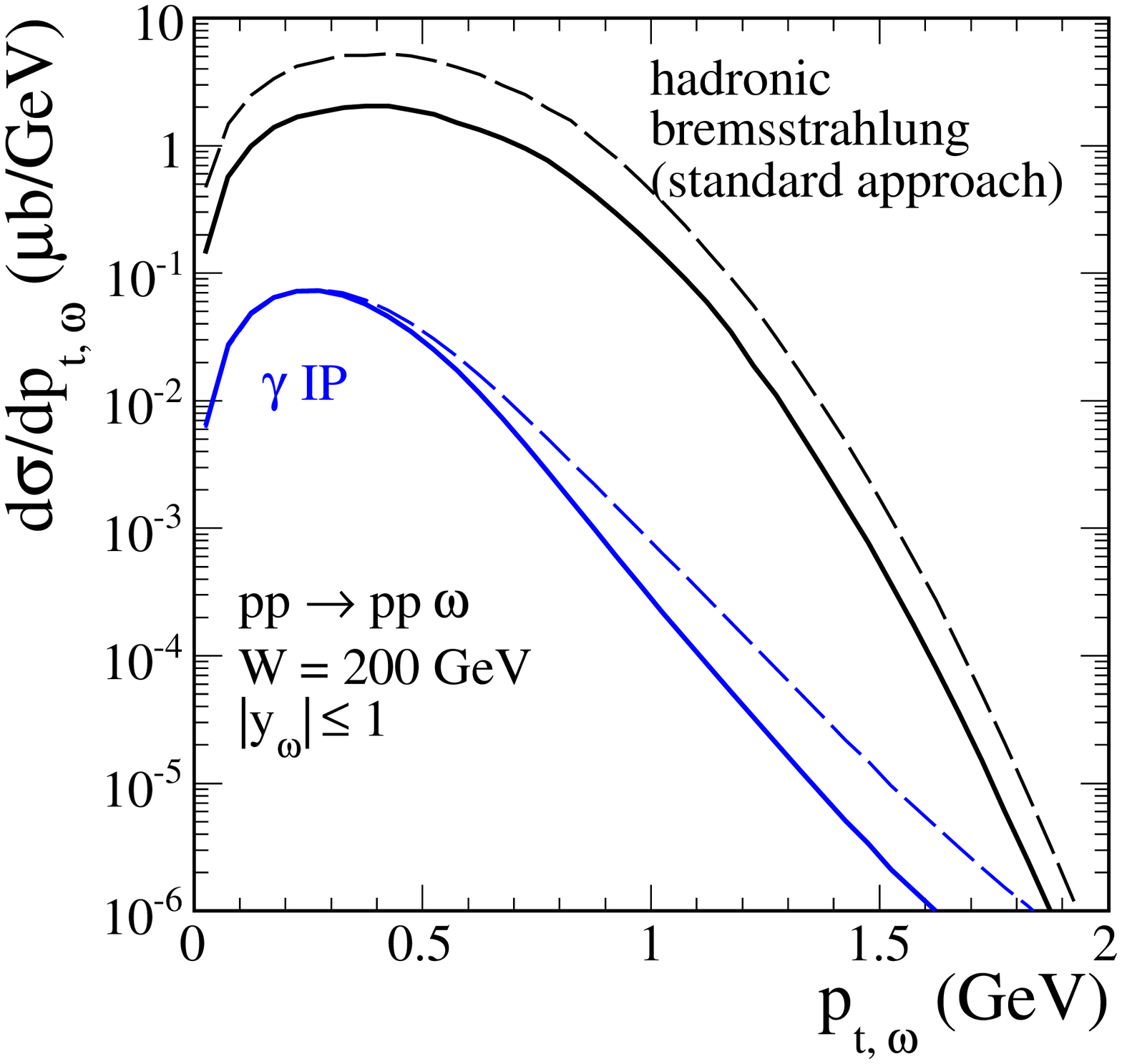}
\includegraphics[width = 0.32\textwidth]{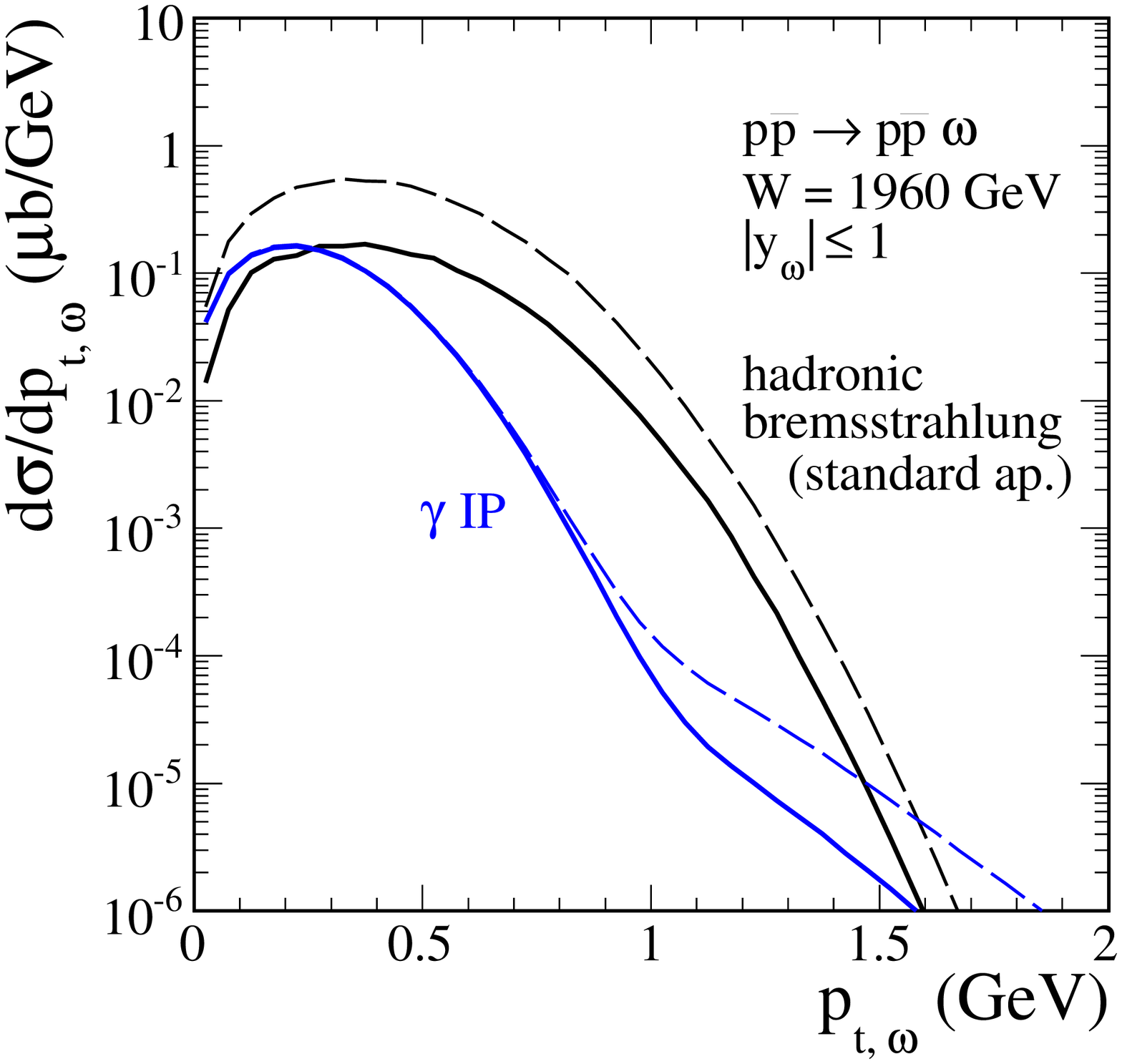}
\includegraphics[width = 0.32\textwidth]{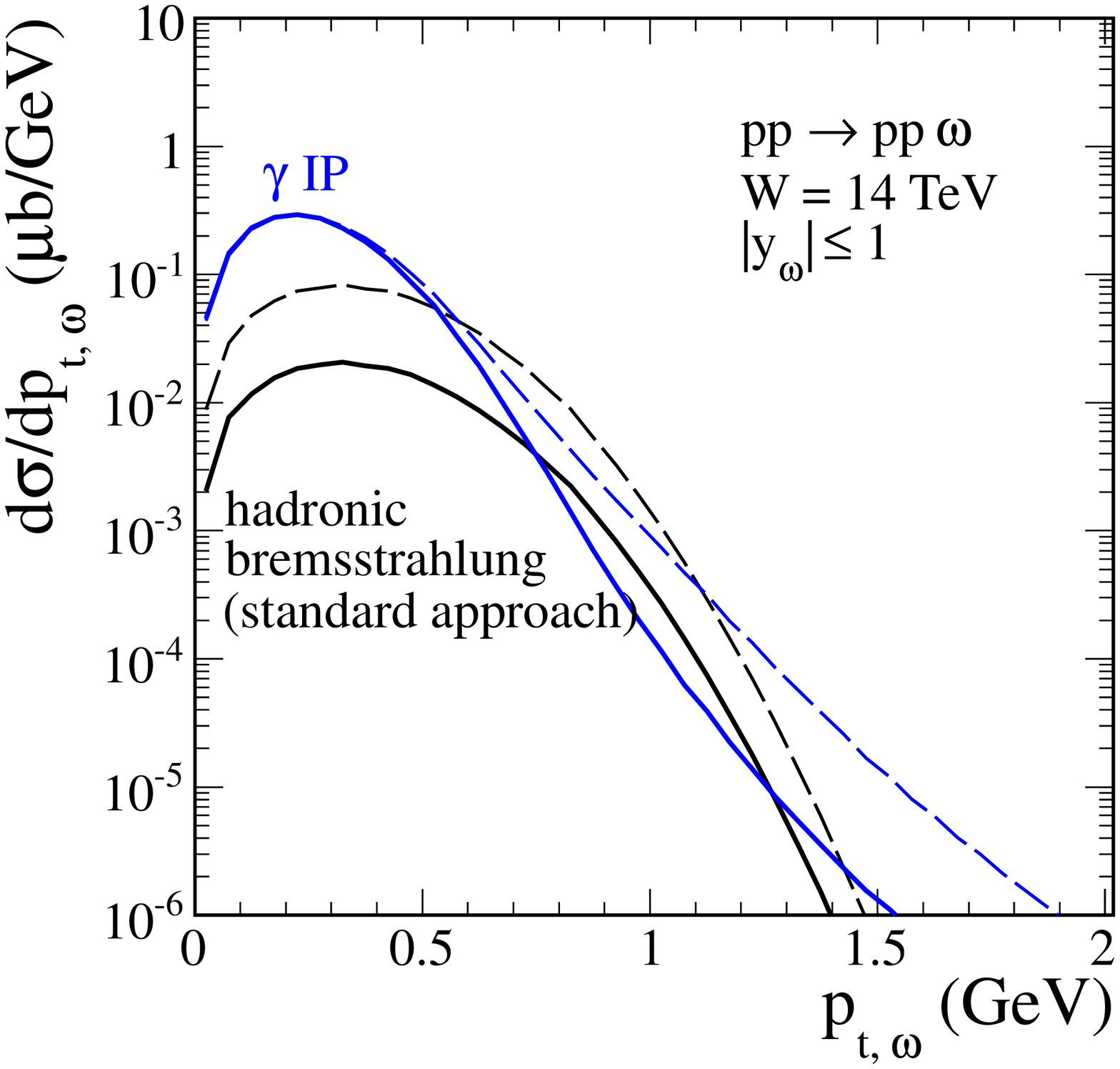}
  \caption{\label{fig:dsig_dpt_limitedy}
  \small
Differential cross section $d\sigma/dp_{t}$
for the $pp(\bar{p}) \to pp(\bar{p}) \omega$ reaction
at $W$ = 200, 1960, 14000 GeV for the limited rapidity range 
-1 $< y_{\omega} <$ 1.
Here the reggeized propagators of omega and standard form factors
are used.
The dashed lines present the contribution without absorption,
while the thick solid lines include the absorption.
}
\end{figure}

\begin{figure}[!h]  
\includegraphics[width = 0.32\textwidth]{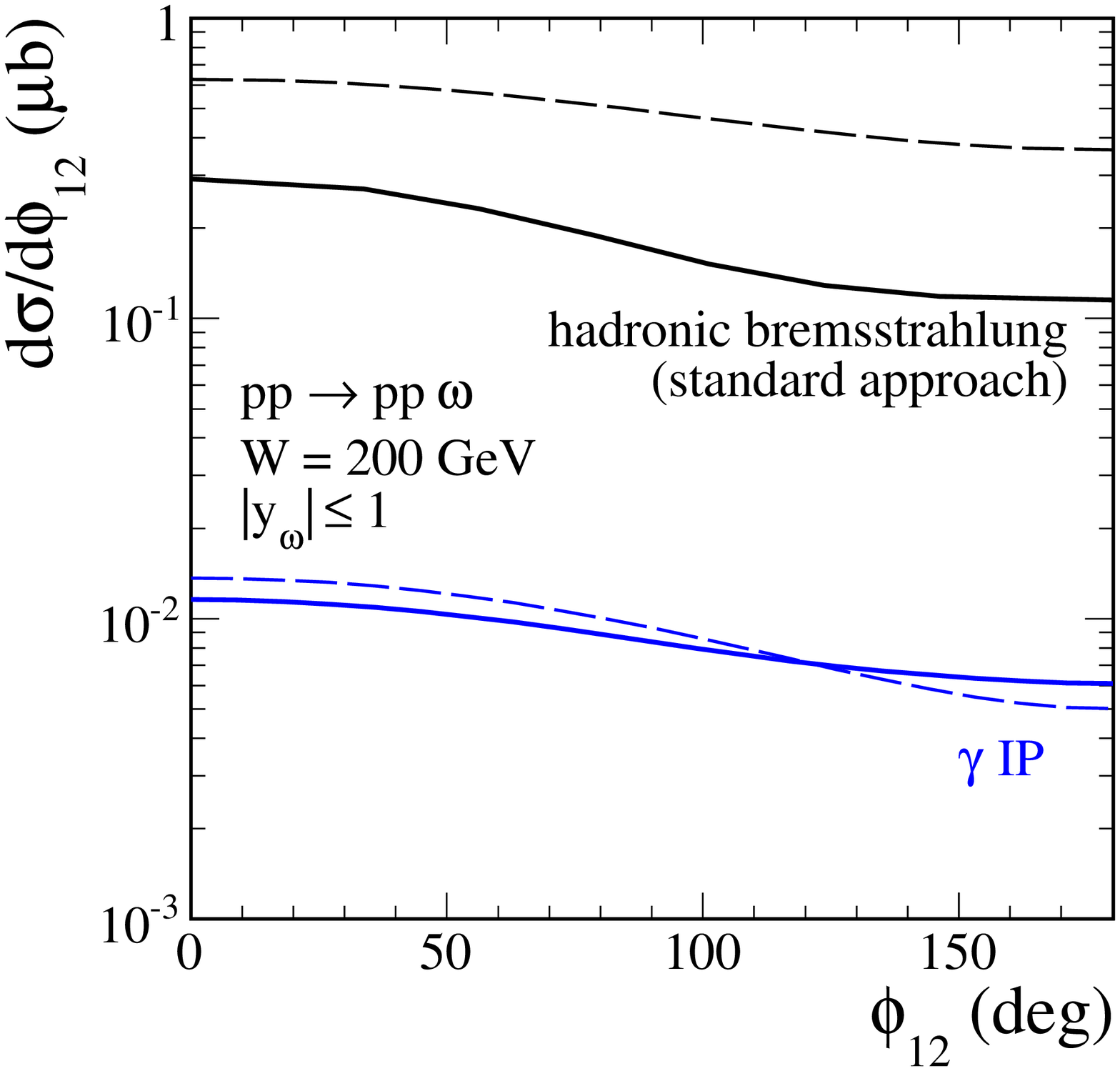}
\includegraphics[width = 0.32\textwidth]{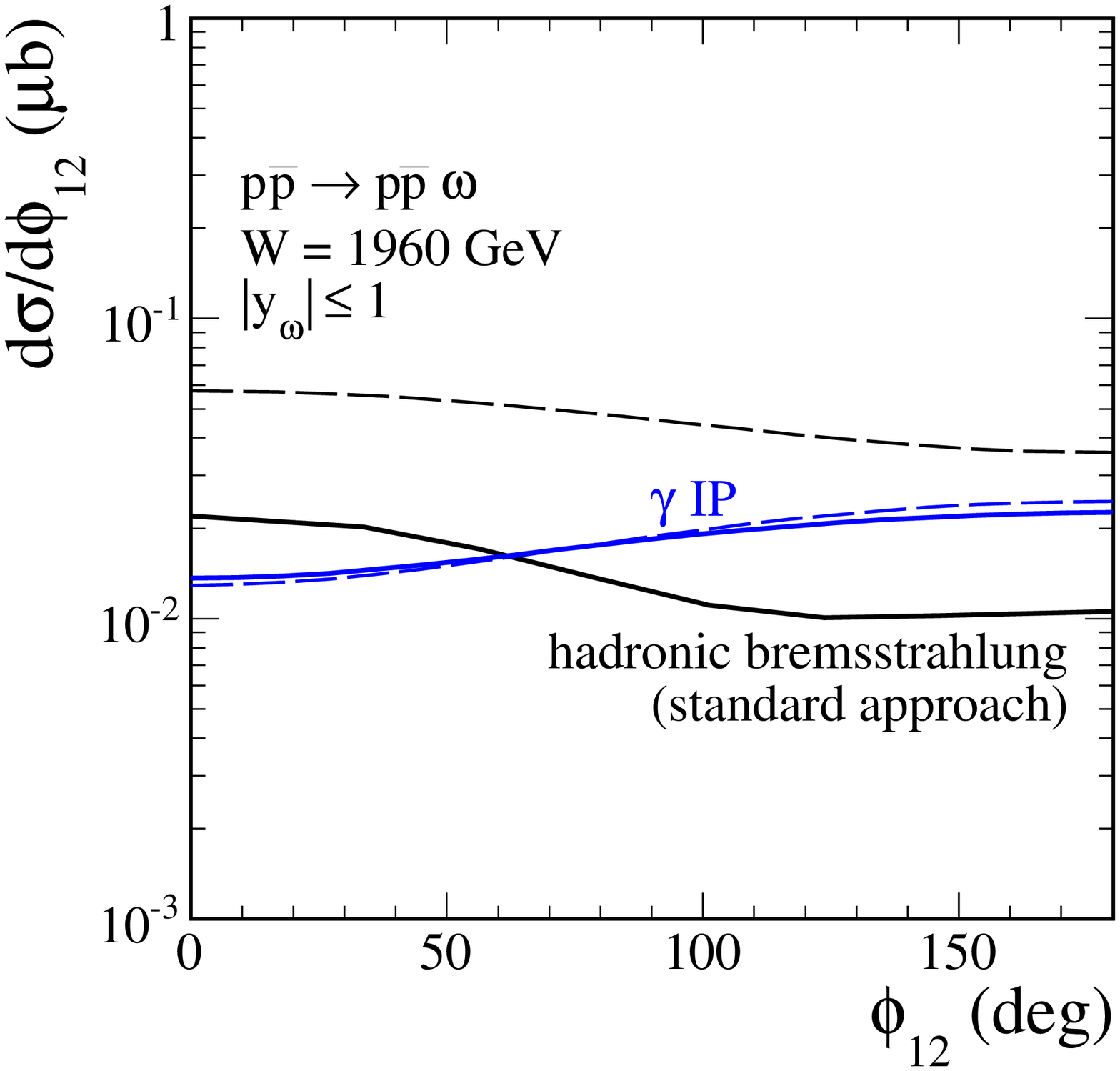}
\includegraphics[width = 0.32\textwidth]{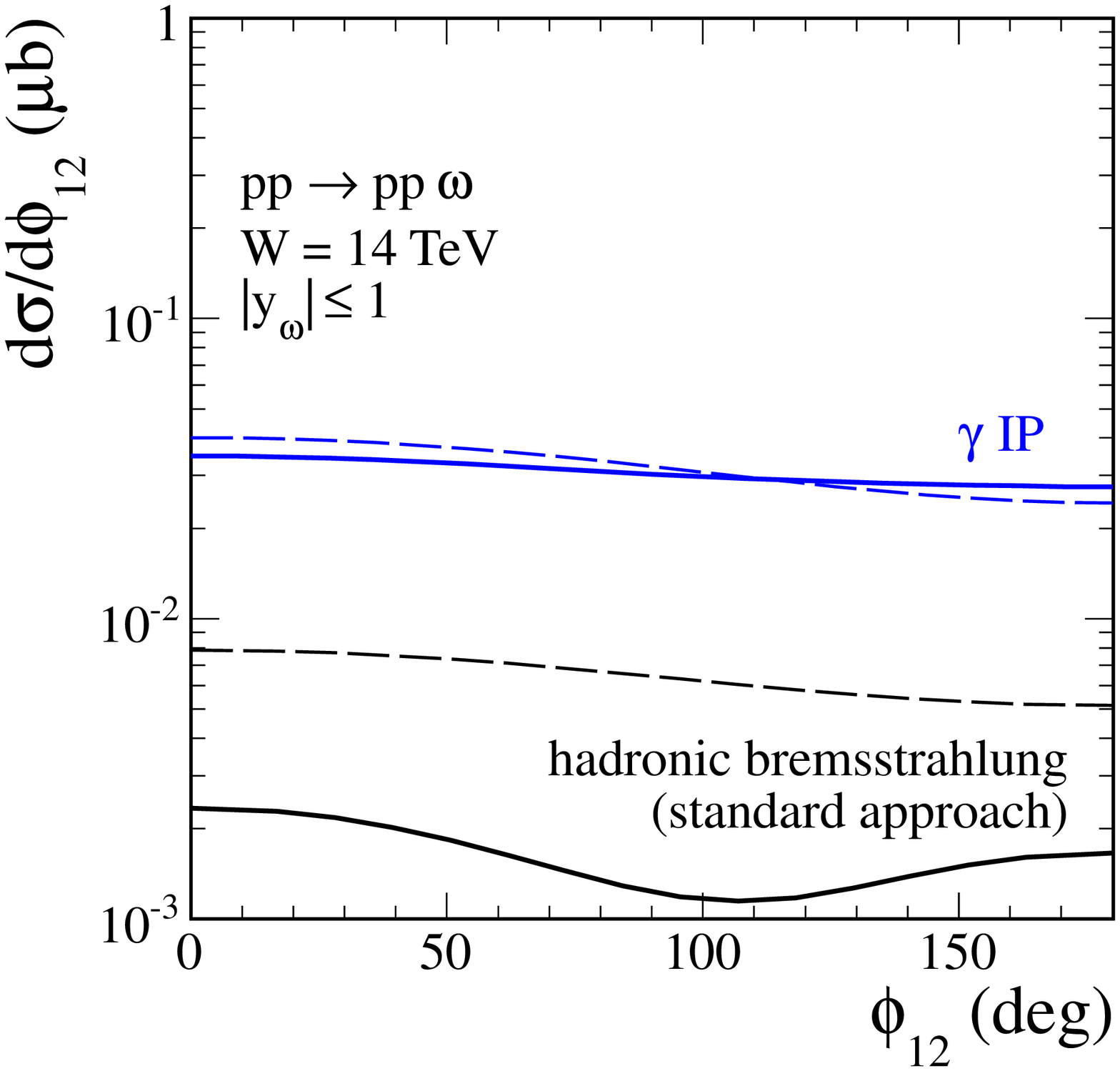}
  \caption{\label{fig:dsig_dphi12_limitedy}
  \small
Differential cross section $d\sigma/d\phi_{12}$
for the $pp(\bar{p}) \to pp(\bar{p}) \omega$ reaction
at $W$ = 200, 1960, 14000 GeV for the limited rapidity range 
-1 $< y_{\omega} <$ 1.
Here the reggeized propagators of omega and standard form factors 
are used.
The dashed lines present the contribution without absorption,
while the thick solid lines include the absorption.
}
\end{figure}

The distributions in the full (pseudo)rapidity range are rather theoretical
and may be difficult to measure.
One may expect that in practice only limited range of (pseudo)rapidity
around $y_{\omega} = 0$ will be available experimentally.
Therefore, as an example, we have made an extra calculation
for a limited rapidity range.
In Fig.\ref{fig:dsig_dpt_limitedy} we show transverse momentum 
distributions for $-1 < y_{\omega} < 1$. Here, as can be seen from
Fig.\ref{fig:dsig_dy}, it is enough to
include only the hadronic bremsstrahlung diagrams e) and f). 
In this case standard form factors are used only.
Please note (see Fig.\ref{fig:dsig_dy}) that in the case of light-cone
form factors the hadronic bremsstrahlung mechanism does not contribute
to the restricted rapidity region. 
For comparison we show the contributions of photoproduction mechanisms
which are calculated fairly precisely as discussed before.
This is very useful in the context of the searches for odderon.

Finally in Fig.\ref{fig:dsig_dphi12_limitedy} 
we show angular correlations between
outgoing protons for $-1 < y_{\omega} < 1$. In the case of light-cone form factors
only the photoproduction mechanism contributes. Testing such distributions
together with rapidity distributions could provide therefore new information
on the mysterious odderon exchange.

\section{Conclusions}

In this paper we have calculated the cross section for $\gamma p \to
\omega p$ reaction at high-energy within a QCD-inspired model.
A good description of the HERA experimental data has been achieved,
comparable as for the $J/\Psi$ and $\phi$
mesons in our previous works.
In the present paper the Gaussian wave function was used 
with parameters adjusted to reproduce
the electronic decay width of $\omega$ meson.

This model is used then to predict the cross sections for the
$p p \to p p \omega$ and $p \bar{p} \to p \bar{p} \omega$ reactions
at high-energies for the first time in the literature.
In contrast to the $J/\Psi$ and $\phi$ exclusive production, 
in the case of the $\omega$ meson different hadronic bremsstrahlung processes 
are possible due to large nonperturbative coupling of the $\omega$ meson 
to the nucleon.
At high energy there is a class of diffractive bremsstrahlung
processes never considered in the literature.

At low energies the hadronic bremsstrahlung contributions dominate over the
photoproduction ones if the standard Mandelstam-dependent form factors are used. 
With increasing energy the hadronic bremsstrahlung
contributions move in rapidity to the fragmentation regions. 
At high energies the photoproduction mechanisms dominate at midrapidities.
We predict a short plateau at midrapidities due to the photoproduction 
mechanism and a significant increase 
towards fragmentation regions (large $|y_{\omega}|$) due to the $\omega$
bremsstrahlung. The identification of 
the increase would be a confirmation of the hadronic bremsstrahlung effects 
discussed here. However, this may be not simple experimentally.
The precisely evaluated photoproduction mechanism constitutes a background for 
the odderon exchange searches.


\vspace{0.5cm}
{\bf Acknowledgments}
This study was partially supported by the Polish grant
of MNiSW N N202 249235 and N N202 322938.

\end{document}